\documentclass[11pt, draftclsnofoot, onecolumn]{IEEEtran}

\usepackage{pifont}
\usepackage{cite}
\usepackage[dvips]{epsfig}
\usepackage{graphicx}
\usepackage{calligra}
\usepackage[T1]{fontenc}
\usepackage{bbold}
\usepackage{mathrsfs}

\usepackage{subfigure}
\usepackage{color}
\usepackage{amsmath}
\usepackage{cases}
 
\usepackage{amssymb}
\usepackage[justification=centering]{caption}
\usepackage{framed}
 
\usepackage{subfigure}
\usepackage{color}
\usepackage{amsmath}
\usepackage{bm}
\usepackage{amssymb}
\usepackage{amsbsy}
\usepackage{bbm}
\usepackage{dsfont}
\usepackage[normalem]{ulem}
\usepackage{xcolor,cancel}
\usepackage{enumerate}
\usepackage{paralist}

\usepackage{algorithm, algorithmic}

\newtheorem{theorem}{Theorem}

\newtheorem{corollary}{Corollary}

\newtheorem{definition}{Definition}

\newtheorem{lemma}{Lemma}

\newtheorem{remark}{Remark}

\newtheorem{assumption}{Assumption}

\newcommand{\bbE}{\mathbbm{E}}

\newcommand{\bbP}{\mathbbm{P}}

\newcommand{\teL}{{\rm{L}}}
\newcommand{\teU}{{\rm{U}}}
\newcommand{\teT}{{\rm{T}}}
\newcommand{\teS}{{\rm{S}}}

\newcommand{\cC}{\mathcal{C}}

\newcommand{\cR}{\mathcal{R}}

\newcommand{\cF}{\mathcal{F}}
\newcommand{\cE}{\mathcal{E}}
\newcommand{\cH}{\mathcal{H}}
\newcommand{\cA}{\mathcal{A}}

\newcommand{\cQ}{\mathcal{Q}}

\newcommand{\cU}{\mathcal{U}}
\newcommand{\cV}{\mathcal{V}}
\newcommand{\cS}{\mathcal{S}}
\newcommand{\cB}{\mathcal{B}}

\newcommand{\bone}{\mathbbm{1}}

\newcommand{\btheta}{{\boldsymbol{\theta}}}

\newcommand{\tu}{\tilde{u}}
\newcommand{\tp}{\tilde{p}}

\newcommand{\tD}{\widetilde{D}}

\newcommand{\hD}{\widehat{D}}

\newcommand{\ignore}[1]{{}}

\newcommand{\lp}{\left(}
\newcommand{\rp}{\right)}

\newcommand{\lsb}{\left[}
\newcommand{\rsb}{\right]}

\newcommand{\lcb}{\left\{}
\newcommand{\rcb}{\right\}}

\newcommand{\lbar}{\left|}
\newcommand{\rclose}{\right.}

\newcounter{parentalgorithm}

\ifCLASSINFOpdf
\else
\fi

\begin{document}
%
\title{Attack Detection in Sensor Network Target Localization Systems with Quantized Data}
%
%
%

\author{{Jiangfan Zhang,~\IEEEmembership{Member,~IEEE}, Xiaodong Wang,~\IEEEmembership{Fellow,~IEEE}, Rick S. Blum,~\IEEEmembership{Fellow,~IEEE}}, and Lance M. Kaplan,~\IEEEmembership{Fellow,~IEEE}


}


\maketitle

\begin{abstract}
We consider a sensor network focused on target localization, where sensors measure the signal strength emitted from the target. Each measurement is quantized to one bit and sent to the fusion center.
A general attack is considered at some sensors that attempts to cause the fusion center to produce an inaccurate estimation of the target location with a large mean-square-error. The attack is a combination of man-in-the-middle, hacking, and spoofing attacks that can effectively change both signals going into and coming out of the sensor nodes in a realistic manner.
We show that the essential effect of attacks is to alter the estimated
distance between the target and each attacked sensor to a different extent, giving rise to a geometric inconsistency among the attacked  and  unattacked sensors. 
 Hence, with the help of two secure sensors, a class of detectors are proposed to detect the attacked sensors by scrutinizing the existence of the geometric inconsistency.
We show that the false alarm and miss probabilities of the proposed detectors decrease exponentially as the number of measurement samples increases, which implies that 
for sufficiently large number of samples, the proposed detectors can identify the attacked and unattacked sensors with any required accuracy. 
\end{abstract}




\section{Introduction}
\label{Section_Introduction}

Sensor networks find wide applications ranging from inexpensive commercial systems to complex military and homeland
defense surveillance systems and have seen ever growing
interest in recent years \cite{akyildiz2002survey}. One important application of sensor networks is to estimate the location of a target in a region of interest (ROI) \cite{kaplan2001maximum, shen2010fundamental1, vempaty2014target}. Recent technological advances in digital wireless communications  and digital electronics have led to the dominance of digital transmission and processing using quantized data in such systems. Hence, a great deal of attention has focused on target localization in sensor networks using quantized data, see \cite{ niu2006target, ozdemir2009channel, vempaty2013localization} for instance.  

Typically, large-scale sensor networks are comprised of low-cost and spatially distributed sensor nodes with limited battery capacity and low computing power, which makes the system vulnerable to cyberattacks by adversaries. This has led to a vast interest in studying the vulnerability of sensor networks in various applications and from different perspectives, see \cite{li2005robust, lee2012characterization, cui2012coordinated, vempaty2013distributed, zhang2015Asymptotically, alnajjab2015attacks, zhang2015distributed, zhang2017functional} and the references therein. Depending on the place where the attack is launched, there are generally three categories of attacks in sensor networks, namely spoofing attacks, hacking attacks, and man-in-the-middle attacks (MiMA). To be specific, the spoofing attack changes the phenomenon observed by the attacked sensors and tampers with the observations coming into the sensors. For example, data-injection attack is one type of spoofing attack \cite{cui2012coordinated}.  The hacking attack aims at hacking into the sensors, modifying the hardware, and/or reprogramming the devices, with the goal of disrupting the data processing in the attacked senors. Note that malfunctions of sensors can also be considered as hacking attacks. The MiMA takes place between the sensors and a fusion center (FC), which maliciously falsifies the data transmitted from the attacked sensors to the FC, see \cite{vempaty2013localization, vempaty2013distributed,zhang2015Asymptotically} for instance.  The main goal of the adversaries is to undermine the sensor network and render the FC to reach an inaccurate estimate of the target location in terms of large mean-square estimation error. A simple and intuitive method to combat the attacks is to identify the attacked sensors so that the FC can either discard data from these sensors, or make use of attacked data to improve its estimate of the target location via jointly estimating the target location and the attacks \cite{vempaty2013distributed, zhang2015Asymptotically, zhang2017functional}.

\subsection{Summary of Results and Main Contributions}

In this paper, we consider a sensor network containing
two widely separated secure sensors which have a very high level of security and thereby are guaranteed to be tamper-proof. The rest of  sensors are unsecure, which are 
subject to arbitrary forms of attacks.  In practice, the two secure sensors can be well protected, built with powerful chips, and supplied with sufficient power, thereby highly sophisticated encryption algorithms and security procedures can be implemented.

This paper aims at developing a general detection approach which does not rely on the form of the attacks or attack parameters, to identify the attacked sensors in the sensor network with provable detection performance guarantee. It is worth mentioning that the problem of attack detection in target localization systems is difficult, since the statistical model of sensor data depend on the target location and the attack strategy which are both unknown to the FC.
By exploring the impact of the attacks on  the statistical model of the sensor data, 
we reveal that the essential effect of attacks is 
to alter the estimated distance between the target and each attacked sensor to a different extent, giving rise to a geometric inconsistency among the attacked and  unattacked sensors. 
Motivated by this fact,  a class of detectors are proposed to detect the attacked sensors via scrutinizing the existence of the geometric inconsistency. To be specific, a naive maximum likelihood estimator (NMLE), the MLE formulated under the assumption of no attack,  is first employed to estimate the  distance between the target and each sensor.  For each unsecure sensor, a circle is generated which is centered at the sensor with radius equal to the NMLE of its distance to the target. For each of the two secure sensors, a ring with some constant width is generated. This ring is centered at the sensor and is bisected by a circle with radius equal to the NMLE of the distance from the sensor to the target. If the  circle of an unsecure sensor passes through the common area of the two rings, the sensor is declared unattacked; otherwise, we declare that it is under attack.
A thorough performance analysis is carried out for the proposed detectors, showing that the false alarm and miss probabilities decrease exponentially  as the number of data samples at each sensor grows, which implies that if 
for a sufficiently large number of samples, the proposed detectors can identify the attacked sensors with an arbitrary level of accuracy.

\subsection{Related Works}

With the proliferation of sensor network applications, there is an increasing concern about the security of sensor networks, see \cite{li2005robust, boukerche2008secure, lee2012characterization, huie2015strategies, capkun2006secure, zhu2011secure} for instance. Most existing works on the security in sensor network target localization systems only consider analog measurements.
However, for a typical sensor network with limited resources, it is desirable that only quantized data is transmitted from sensors to the FC \cite{niu2006target, ozdemir2009channel, vempaty2013localization}. Moreover, there is a lack of theoretical performance analysis of  attack detection strategies.

Attack detection in the context of target localization with quantized data has not been well investigated in the literature. In \cite{vempaty2013localization}, a specific attack model is considered and a practical approach is proposed to detect attacks in target localization systems. In particular,
several secure sensors are employed to provide a coarse estimate of the target location, and then the expected behaviors of attacked and unattacked sensors are calculated based on the coarse estimate and the attack model.
This method is based on heuristic and there is no detection performance guarantee. In our proposed approach, the estimate of the target location is not required, and moreover, the attack detection performance  is rigorously investigated, which demonstrates that any identification accuracy can be achieved 
if the number of  data samples is sufficiently large.  In addition, the  approach  in  \cite{vempaty2013localization} requires the knowledge of the statistical model of the attack, which is not required by our proposed approach.

The remainder of the paper is organized as follows. Section \ref{Section_Problem}  describes the system and adversary model. In Section \ref{Section_Attack_Detection}, a class of detectors are proposed to identify the attacked sensors in the sensor network. Section \ref{Section_Performance_Analysis} investigates the performance of the proposed detectors.
In Section \ref{Section_Numerical}, several numerical results are provided to corroborate our theoretical analysis. Finally, Section \ref{Section_Conclusion} provides our conclusions.

\section{System and Adversary Models}
\label{Section_Problem}

In this section, the system and general attack models are introduced. We also demonstrate how the general attack model relates to some popular  forms of attacks in practice.

\subsection{System Model}

Consider a sensor network consisting of $N$ sensors and a FC to estimate the location of a target at ${{\btheta}_\teT} = [{{x_\teT},{y_\teT}} ]$, where $x_\teT$ and $y_\teT$ denote the coordinates of the target location on the two-dimensional plane. For the $j$-th sensor, we use $\btheta_j = [{{x_j},{y_j}} ]$ to denote its location. Besides the $N$ sensors, there also exist two secure sensors in the sensor network which are labeled as the $(N+1)$-th and $(N+2)$-th sensors, respectively. These two secure sensors are well protected and thereby are guaranteed to be tamper proof, while the other $N$ sensors are unsecure, which are subject to threat from adversaries.
We assume that the signal radiated from the target obeys an isotropic power attenuation model, and each sensor observes $K$ data samples. The $k$-th data sample at the $j$-th sensor is described as
\begin{equation} \label{signal_model}
{s_{jk}} = {P_0}{\left( {\frac{{{D_0}}}{{{D_j}}}} \right)^\gamma } + {n_{jk}}, \;   j=1,2,...,N+2,
\end{equation}
where the distance ${D_j}$ between the $j$-th sensor and the target is defined by
\begin{equation}
{D_j} \buildrel \Delta \over = \left\| {{{\btheta}_j} - {{\btheta}_\teT}} \right\|= \sqrt {{{\left( {{x_j} - {x_\teT}} \right)}^2} + {{\left( {{y_j} - {y_\teT}} \right)}^2}}, \;  j=1,2,...,N+2,
\end{equation}
the quantity $P_0$ is the power  measured at a reference distance $D_0$, $\gamma $ is the path-loss exponent, 
and ${n_{jk}}$ denotes the additive noise sample with 
probability density function (pdf) $f_j({n_{jk}})$. 

We assume that $P_0$, $D_0$, $\gamma$, $\{f_j( \cdot )\}_{j=1}^{N+2}$, and $\{{\btheta}_j\}_{j=1}^{N+2}$ are known to the FC. Moreover, we assume $\{ {{n_{jk}}} \}$ are independent, and for each $j$, $\{ {{n_{jk}}} \}_{k=1}^K$ is an identically distributed sequence. In addition, we assume that the target stays in a specified ROI ${\cA}$ where no sensor exists. By defining
\begin{align}
{D_{\teL}} & \buildrel \Delta \over = \mathop {\min }\limits_{j=1,2,...,N+2} \mathop {\inf }\limits_{{\btheta} \in {\cA}} \left\| {{{\btheta}_j} - {\btheta}} \right\| >0, \\
\text{and} \quad {D_{\teU}} & \buildrel \Delta \over = \mathop {\max }\limits_{j=1,2,...,N+2} \mathop {\sup }\limits_{{\btheta} \in {\cA}} \left\| {{{\btheta}_j} - {\btheta}} \right\| < \infty,
\end{align}
we know that for any $j \in \{1,2,...,N+2\}$,
\begin{equation} \label{D_j_range}
{D_j} \in \left[ {  D_{\teL}, D_{\teU} } \right].
\end{equation}
Regarding the secure sensors and the ROI ${\cA}$, we make the following assumption.
\begin{assumption} \label{Assumption_Area}
	The secure sensors are widely separated so that
	\begin{equation} \label{assumption_Area_Widely_Separated}
	D_\teS \triangleq \left\| {{{\btheta}_{N+1}} - {{\btheta}_{N+2}}} \right\| > D_\teU - D_\teL + 2\Upsilon_1
	\end{equation}
	for some positive constant $\Upsilon_1$. In addition, the ROI ${\cA}$ is contained in one of the two half spaces produced by dividing the whole space  by the line passing through the two secure sensors.
	By the triangle inequality of sides, we assume
	\begin{equation} \label{assumption_D_N_1_N_2}
	\mathop {\inf }\limits_{{\btheta_{\rm{T}}} \in {\cA}} \{ D_{N+1} + D_{N+2}\} > D_\teS + 2\Upsilon_2
	\end{equation}
	for some positive constant $\Upsilon_2$.
\end{assumption}

Due to the 
low-rate communication constraint between the sensors and the FC, each sensor $j$ quantizes its sample $s_{jk}$ to one bit and then transmits the bit to the FC. 
For simplicity, we assume that the sensors employ the following threshold quantizers $\{\cQ_j\}_{j=1}^{N+2}$
\begin{equation} \label{binary_quantizer}
{u_{jk}} =  \cQ_j \lp s_{jk} \rp \triangleq \mathbbm{1}\left\{ {{s_{jk}} \in \left( {{\tau}_j ,\infty } \right)} \right\}, \;  j=1,2,...,N+2, \;  k=1,2,...,K,
\end{equation}
where ${\tau}_j$ is the threshold employed at the $j$-th sensor and we assume that 
the thresholds $\{ {\tau}_j \}_{j=1}^{N+2}$ are known to the FC. 


Using (\ref{signal_model}) and (\ref{binary_quantizer}),
define
\begin{equation} \label{Define_p_j}
p_j(\btheta_\teT) \triangleq \Pr \left( {{u_{jk}} = 0\left| {\btheta}_\teT  \right.} \right) = F_j\left( {\tau_j  - {P_0}{\left( {\frac{{{D_0}}}{{D_j}}} \right)^\gamma } } \right),
\end{equation}
%
where $F_j\left( x \right){\rm{ }} \buildrel \Delta \over = \int_{ - \infty }^x {f_j\left( t \right)dt} $.
By employing (\ref{D_j_range}) and (\ref{Define_p_j}), we can define
\begin{align} \label{Definition_rho_L}
\rho_{j}^{(\teL)} & \triangleq  \mathop {\inf }\limits_{{\btheta} \in {\cA}} p_j(\btheta) =  F_j\left( {\tau_j  - {P_0}{\left( {\frac{{{D_0}}}{{D_\teL}}} \right)^\gamma } } \right),  \\ \label{Definition_rho_U}
\rho_{j}^{(\teU)} & \triangleq \mathop {\sup }\limits_{{\btheta} \in {\cA}} p_j(\btheta) =  F_j\left( {\tau_j  - {P_0}{\left( {\frac{{{D_0}}}{{D_\teU}}} \right)^\gamma } } \right),
\end{align}
and hence,
\begin{equation} \label{p_j_range}
p_j(\btheta_\teT) \in \lsb \rho_{j}^{(\teL)}, \rho_{j}^{(\teU)} \rsb, \;  j=1,2,...,N+2.
\end{equation}

We assume that $f_j(x)$ is continuous, and ${F_j^{ - 1}}\left( x \right)$ exists and is differentiable over the open interval $\left( {0,1} \right)$ for each $j$. Noticing that $\frac{{\partial F_j^{ - 1}\left( x \right)}}{{\partial x}} = [{{{f_j}( {F_j^{ - 1}( x )} )}}]^{-1}$, the differentiability of  ${F_j^{ - 1}}\left( x \right)$ implies  $0<f_j(x)<\infty$ over $\{x | F_j(x) \in (0,1) \}$, and therefore, $F_j(x)$ is strictly increasing over $\{x | F_j(x) \in (0,1) \}$.

It is clear that if there exists some $\btheta \in {\cal A}$ such that
\begin{equation}
\tau_j - {P_0}{\left( {\frac{{{D_0}}}{{\| \btheta_j - \btheta  \|}}} \right)^\gamma } \notin {\rm{supp}} \lp f_j \rp \triangleq \lcb x | f_j(x) \ne 0 \rcb,
\end{equation}
then $F_j( {\tau_j  - {P_0}{( {\frac{{{D_0}}}{{\| \btheta_j - \btheta  \|}}} )^\gamma } } ) =0 \text{ or } 1$, and hence, the quantized data from the $j$-th sensor is useless in estimating $\btheta$. To this end, we assume that the quantizers are well designed, and thereby $\tau_j$, $D_\teL$ and $D_\teU$ satisfy
\begin{equation} \label{D_L_D_U_support}
\inf \lcb {\rm{supp}} \lp f_j \rp \rcb < \tau_j  - {P_0}{\left( {\frac{{{D_0}}}{{D_\teU}}} \right)^\gamma }  <   \tau_j  - {P_0}{\left( {\frac{{{D_0}}}{{D_\teL}}} \right)^\gamma } < \sup \lcb {\rm{supp}} \lp f_j \rp  \rcb,
\end{equation}
which yields
\begin{equation} \label{F_j_range}
0 < {F_j}\left( {{\tau _j} - {P_0}{{\left( {\frac{{{D_0}}}{{{D_{\rm{L}}}}}} \right)}^\gamma }} \right) < {F_j}\left( {{\tau _j} - {P_0}{{\left( {\frac{{{D_0}}}{{{D_{\rm{U}}}}}} \right)}^\gamma }} \right) < F_j\lp \tau_j \rp \le 1, \;   j=1,2,...,N+2,
\end{equation}
since $F_j(\cdot)$ is strictly increasing,  from (\ref{Definition_rho_L}) and (\ref{Definition_rho_U}), we know 
\begin{equation} \label{rho_L_rho_U_less_1}
0 < \rho_{j}^{(\rm{L})} < \rho_{j}^{(\rm{U})} < F_j\lp \tau_j \rp \le 1.
\end{equation}

\subsection{Adversary Model}

We consider a general attack model which brings about a change in the statistical model of $u_{jk}$.
Let $\cU$ and $\cV$  denote the set of unattacked and attacked sensors, respectively.

	\begin{figure}[htb]
		\centerline{
			\includegraphics[width=0.5\textwidth]{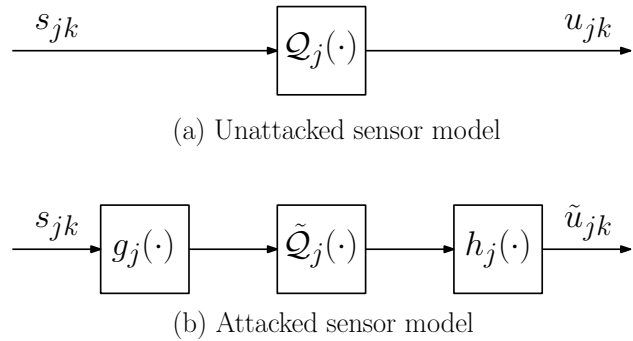}
		}
		\caption{Unattacked and attacked sensor models.
		}	
		\label{Fig_Attack_Model}	
	\end{figure}

In general, if $j \in \cV$,  three types of possible attacks can affect the $j$-th sensor, which are illustrated in Fig. \ref{Fig_Attack_Model} (b). First, the adversaries can 
tamper with the observations $\{s_{jk}\}_{k=1}^{K}$. Such attacks are called spoofing attacks, which can be represented by a mapping $g_j(\cdot)$. 
The second type of attack which we call hacking, aims at modifying the sensor hardware and/or software, and thereby modifying the quantizer $\cQ_j(\cdot)$ to ${\tilde{\cQ}}_j(\cdot)$ in the attacked sensors as shown in Fig. \ref{Fig_Attack_Model} (b).
The last type of possible attack occurs between the sensors and the FC, which is referred to as man-in-the-middle attacks (MiMA). The MiMA 
can be described by a mapping $h_j(\cdot)$ that modifies the quantized data before it arrives at the FC. Therefore, the post attack quantized data 
can be generally expressed as
\begin{equation}
\tu_{jk} = h_j \lp {\tilde{\cQ}}_j \lp  g_j\lp s_{jk} \rp  \rp \rp.
\end{equation}

With regard to the alphabet set of $\tu_{jk}$, we make the following assumption.
\begin{assumption} \label{Assumption_tu_range}
We assume that if $j\in \cV$, then the alphabet set of $\tu_{jk}$ is still $\{0,1\}$. Otherwise, the detection of attacks is trivial.
\end{assumption}


Define
\begin{equation} \label{tp}
\tp_j(\btheta_\teT) \triangleq \Pr \left( {{\tu_{jk}} = 0\left| {\btheta}_\teT  \right.} \right) = p_j({\btheta_\teT}) + \Psi_j, \;  j=1,2,...,N,
\end{equation}
where the quantity $\Psi_j$ represents the impact of the attacks on the statistical model of the data. Clearly, if $\Psi_j=0$,
then we can ignore the corresponding attack, since it is ineffective from the perspective of the FC. Hence, without loss of generality, if $j \in \cV$, then we assume $\Psi_j \ne 0$, while if $j \in \cU$, then $\Psi_j = 0$.


To illustrate (\ref{tp}) in a concrete way, 
we take the MiMA as an example.
Under a class of MiMAs \cite{vempaty2013localization, vempaty2013distributed, zhang2015Asymptotically}, 
the quantized data $u_{jk}$ is flipped with probability $\psi_{j,i}$ if $u_{jk}=i$ for $i\in\{0,1\}$, i.e., if the $j$-th sensor is attacked, 
\begin{equation} \label{man_in_the_middle_attack_Rule}
\lcb \begin{split}
\Pr\lp \tu_{jk} = 1 \lbar u_{jk} = 0 \rclose \rp =  \psi_{j,0}, \\
\Pr \lp \tu_{jk} = 0 \lbar u_{jk} = 1 \rclose \rp =  \psi_{j,1}, 
\end{split} \rclose
\end{equation}
where  $\psi_{j,i} \in [0,1]$. 
Using (\ref{man_in_the_middle_attack_Rule}), we have
\begin{align} \label{tp_example}
 \tp_j(\btheta_\teT)  & = \lp 1 - \psi_{j,0} - \psi_{j,1} \rp p_j(\btheta_\teT) + \psi_{j,0}, \\\label{delta_example}
{\text{and}} \qquad \Psi_j  &   = \psi_{j,0} -  \lp \psi_{j,0} + \psi_{j,1} \rp  p_j(\btheta_\teT).
\end{align}

Besides the man-in-the-middle attacks, the spoofing attacks  can also be shown to agree with  (\ref{tp}) \cite{li2005robust, lee2012characterization, zhang2017functional}.


From a practical point of view, the following assumptions on the attacks are made throughout this paper.
\begin{assumption} \label{Assumption_Attack}
	$\ $
	\begin{enumerate}
		\item \emph{Subtle Attacks.} 
By the strong law of large numbers, we know that as $K \to \infty$, $\frac{1}{K}\sum_{k = 1}^K \lp 1 - {{\tu_{jk}}} \rp \to  \tp_j(\btheta_\teT)$ almost surely. Thus, if $\tp_j(\btheta_\teT) \notin [\rho_{j}^{(\teL)}, \rho_{j}^{(\teU)}]$, then with sufficient observations, the attack against the $j$-th sensor can be detected at the FC by checking whether $\frac{1}{K}\sum_{k = 1}^K \lp 1 - {{\tu_{jk}}} \rp $ is in the range $[\rho_{j}^{(\teL)}, \rho_{j}^{(\teU)}]$. For this reason, in order to reduce the possibility of being detected, the adversaries should ensure 
		\begin{equation}  \label{tp_j_range}
		\tp_j(\btheta_\teT) \in \lsb \rho_{j}^{(\teL)}, \rho_{j}^{(\teU)} \rsb, \;   j \in \cV.
		\end{equation}
		\item \emph{Significant Attacks.} In order to bring about sufficient impact on the statistical characterization of the bits from the attacked sensors, every adversary is required to guarantee a minimum distortion, i.e.,
		\begin{equation} \label{Psi_j_larger}
		| \Psi_j | > \kappa, \;   j \in {\cal V},
		\end{equation}
		for some positive constant $\kappa$. Otherwise, the attacks can be ignored.
	\end{enumerate}
\end{assumption}


Our problem is to design an efficient strategy for the FC to identify the attacked sensors, based on the binary observations it receives from all sensors, and to provide a performance analysis on the proposed attack detection strategy.

\section{Attack Detectors Based on Naive Maximum Likelihood Estimator}
\label{Section_Attack_Detection}

In this section,  we first show that by employing a naive maximum likelihood estimator (NMLE), a geometric inconsistency among each attacked sensor and other unattacked sensors can be utilized to distinguish between the attacked and unattacked ones. Then, a class of detectors which are based on the  NMLE are proposed to detect the attacks in the sensor network.


\subsection{Naive Maximum Likelihood Estimator and Geometric Inconsistency}

For any $j$, from (\ref{Define_p_j}) and by employing the existence of ${F_j^{ - 1}}\left( x \right)$, we can obtain
\begin{equation} \label{D_j}
D_j  = D_0 P_0^{\frac{1}{\gamma }}{\left[ {{\tau _j} - {F_j^{ - 1}}\left(  p_j(\btheta_\teT) \right)} \right]^{ - \frac{1}{\gamma } }}.
\end{equation}


Then the NMLE, which is the MLE under the assumption of no attack, of $D_j$ is given by
\begin{equation} \label{Definition_hD_jK}
{{\widehat D}_{j}^{(K)}} = {D_0}P_0^{\frac{1}{\gamma }}{\left[ {{\tau _j} - F_j^{ - 1}\left( { \xi_{j}^{(K)} } \right)} \right]^{ - \frac{1}{\gamma }}},
\end{equation} 
\begin{equation} \label{Definition_xi_jK}
{\text{where}} \quad \xi_{j}^{(K)} \triangleq \frac{1}{K}\sum\limits_{k = 1}^K \lp 1 - {{\tu_{jk}}} \rp.
\end{equation}

\begin{figure}
	\begin{minipage}{0.5\linewidth}
		\centering
		\includegraphics[width=0.68\textwidth]{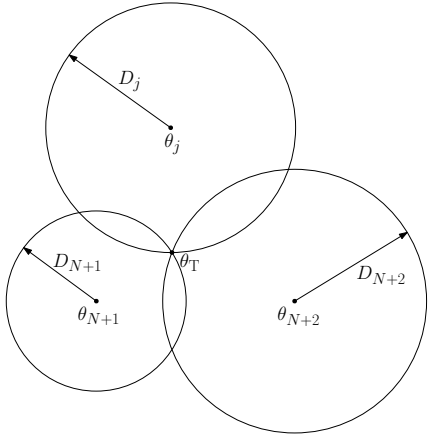}
		\caption{Geometric consistency among the $j$-th, $(N+1)$-th and $(N+2)$-th sensors  when $j\in \cU$.
		}	
		\label{Fig_Geometric_consistency}	
	\end{minipage}%
	\begin{minipage}{0.5\linewidth}
		\centering
		\includegraphics[width=0.68\textwidth]{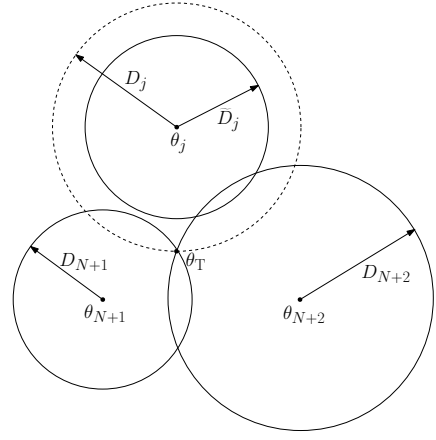}
		\caption{Geometric inconsistency among the $j$-th, $(N+1)$-th and $(N+2)$-th sensors when $j\in \cV$.
		}	
		\label{Fig_Geometric_inconsistency}	
	\end{minipage}
\end{figure}

Furthermore, define
\begin{equation} \label{Define_tilde_D_j}
\tD_j \triangleq  {D_0}P_0^{\frac{1}{\gamma }}{\left[ {{\tau _j} - F_j^{ - 1}\left( \tp_j \lp \btheta_\teT \rp \right)} \right]^{ - \frac{1}{\gamma }}}.
\end{equation}
It is seen from (\ref{Define_tilde_D_j}) that $\tD_j$ is a monotonic function of $\tp_j \lp \btheta_\teT \rp$, and since from (\ref{Psi_j_larger}), we know $\tp_j \lp \btheta_\teT \rp \ne p_j \lp \btheta_\teT \rp$, we have $\tD_j \ne D_j$. What's more,
by the strong law of large numbers, we know
\begin{equation} \label{NMLE_D_j}
{\widehat{D}}_j^{(K)} \to \left\{ \begin{array}{l}
{D_j},  \quad \text{if} \quad   j \in \cU       \\
{{\widetilde D}_j}, \quad \text{if} \quad   j \in \cV
\end{array} \right. \text{ almost surely}, \; \text{as} \;  K \to \infty.
\end{equation}
This implies that, from the perspective of the NMLE, if $j \in \cV$, the essential effect of the attack is a falsification of the distance $D_j$ between the target and the $j$-th sensor to some different $\tD_j$. 
This gives rise to a geometric inconsistency between the $j$-th sensor and the two secure sensors, which is illustrated in terms of the difference between Fig. \ref{Fig_Geometric_consistency} and Fig. \ref{Fig_Geometric_inconsistency}. Specifically,  if $j \in \cU$, as illustrated in  Fig. \ref{Fig_Geometric_consistency}, the three circles centered at the $j$-th, $(N+1)$-th and $(N+2)$-th sensors and with radii $D_j$, $D_{N+1}$ and $D_{N+2}$, respectively, intersect at the point $\btheta_{\teT}$; while if $j \in \cV$, then the three circles do not intersect at $\btheta_{\teT}$ as illustrated in  Fig. \ref{Fig_Geometric_inconsistency}.

Motivated by this fact, consider three circles centered at the $j$-th, $(N+1)$-th and $(N+2)$-th sensors and with radii $\hD_j^{(K)}$, $\hD_{N+1}^{(K)}$ and $\hD_{N+2}^{(K)}$, respectively. If $j \in \cV$, then from (\ref{NMLE_D_j}), we know that with sufficiently large $K$ and Assumption \ref{Assumption_Attack}, it is impossible for these three circles to intersect at a common point. 
This observation forms the basis of the proposed attack detection strategy. 

\subsection{Attack Detection Strategy}

\begin{figure}
	\begin{minipage}{0.33\linewidth}
		\centering
		\includegraphics[width=0.68\textwidth]{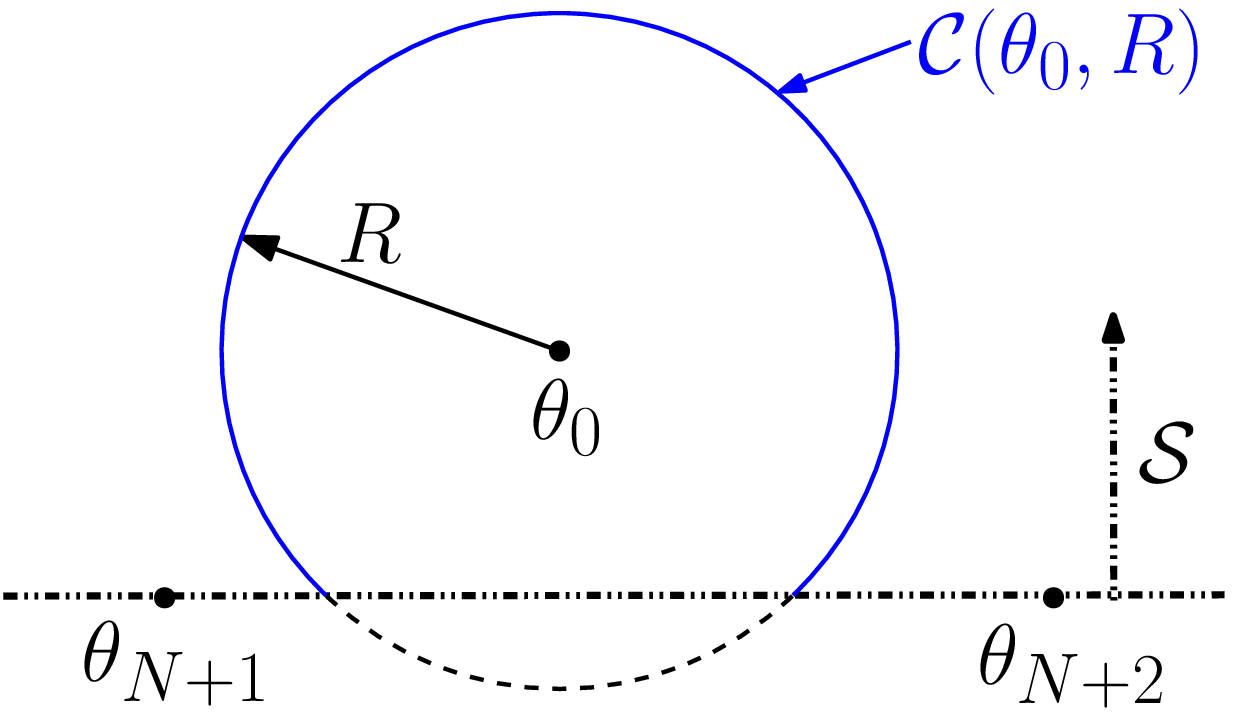}
		\caption{Geometric illustration of $\cC(\btheta_0, R)$.
		}	
		\label{Fig_Circle}	
	\end{minipage}%
	\begin{minipage}{0.33\linewidth}
		\centering
		\includegraphics[width=0.68\textwidth]{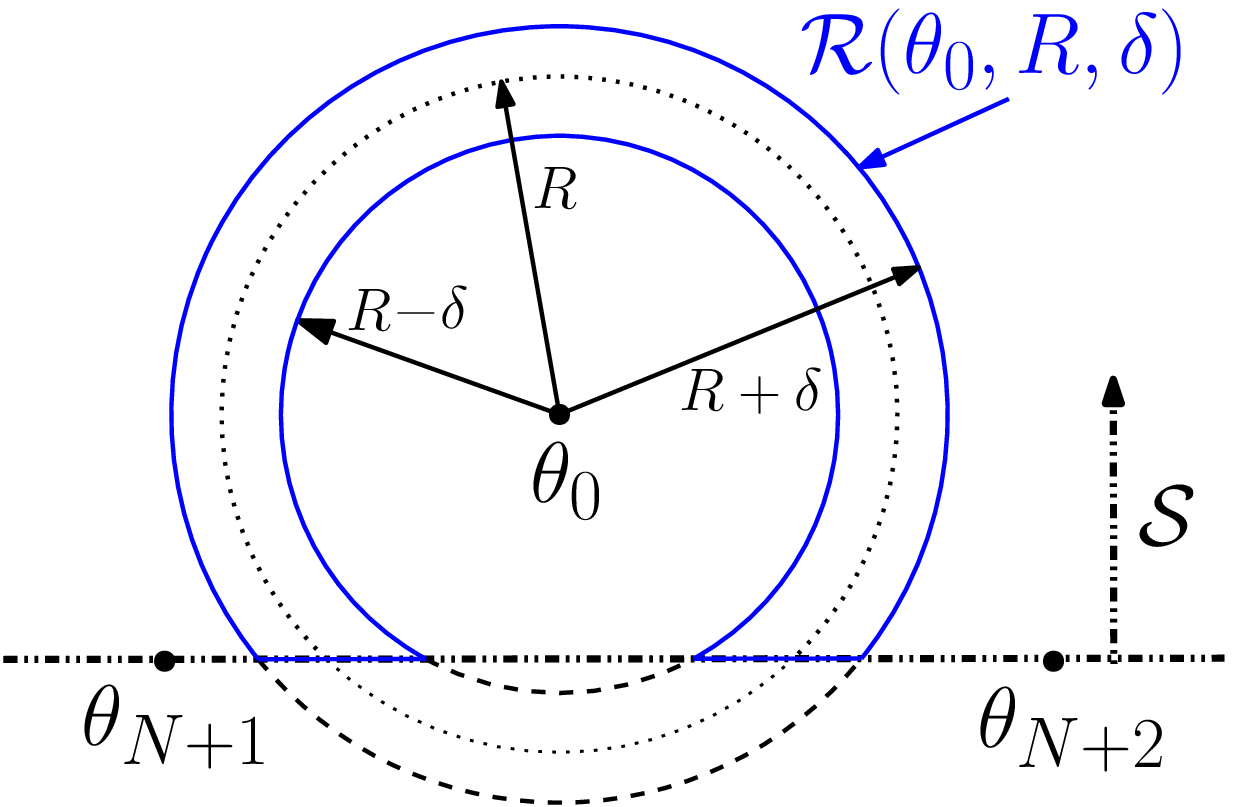}
		\caption{Geometric illustration of $\cR(\btheta_0, R, \delta)$.
		}	
		\label{Fig_Ring}	
	\end{minipage}
	\begin{minipage}{0.33\linewidth}
		\centering
		\includegraphics[width=0.68\textwidth]{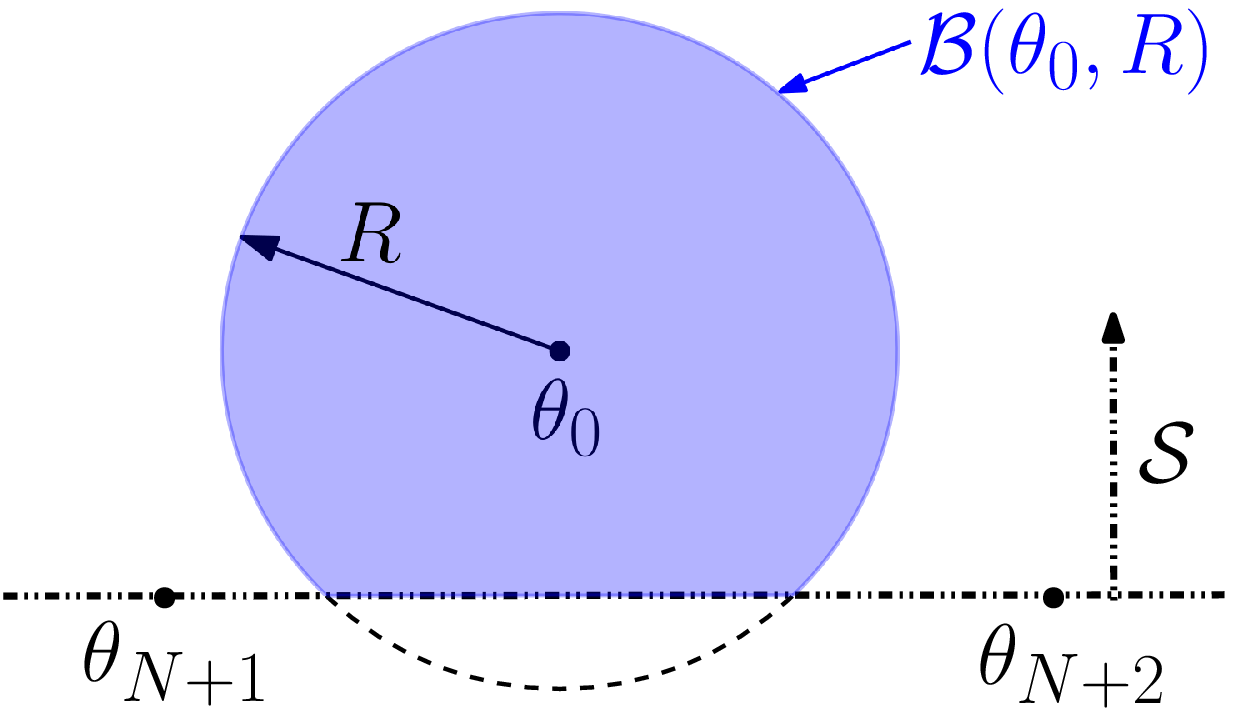}
		\caption{Geometric illustration of $\cB(\btheta_0, R)$.
		}	
		\label{Fig_Ball}	
	\end{minipage}
\end{figure}

In order to mathematically formulate the attack detector, 
we first define three geometric shapes.  According to Assumption \ref{Assumption_Area}, the ROI ${\cA}$ is contained in one of the two half spaces produced by dividing the whole space by the line passing through the two secure sensors. We use $\cS$ to represent this half space.
Let $\cC(\btheta_0, R)$ denote the intersection of $\cal S$ and the circle centered at $\btheta_0$ and with radius $R$, i.e.,
\begin{equation} \label{Definition_circle}
\cC(\btheta_0, R) \triangleq \lcb \btheta \in \cS \left| \| \btheta - \btheta_0 \| = R \right. \rcb,
\end{equation}
which is illustrated by the blue curve in Fig. \ref{Fig_Circle}.
Let $\cR(\btheta_0, R, \delta)$ denote the intersection of $\cal S$ and the ring centered at $\btheta_0$, with radius $R$ and width $\delta$, i.e.,
\begin{equation} \label{Definition_ring}
\cR(\btheta_0, R, \delta) \triangleq \lcb \btheta \in \cS \left|    R -\delta\le \| \btheta - \btheta_0 \| \le R +\delta \right. \rcb.
\end{equation}
The region enclosed by the blue boundary in Fig. \ref{Fig_Ring} depicts an example of $\cR(\btheta_0, R, \delta)$.
Let $\cB(\btheta_0, R)$ denote the intersection of  $\cal S$ and the ball centered at $\btheta_0$ and with radius $R$, i.e.,
\begin{equation} \label{Definition_ball}
\cB(\btheta_0, R) \triangleq \lcb \btheta \in \cS \left|   \| \btheta - \btheta_0 \| \le R  \right. \rcb.
\end{equation}
which is the blue region in Fig. \ref{Fig_Ball}.

It is worth mentioning that  even though $j \in \cU$, due to the estimation error with finite $K$, 
the three circles centered at the $j$-th, $(N+1)$-th and $(N+2)$-th sensors and with radii $\hD_j^{(K)}$, $\hD_{N+1}^{(K)}$ and $\hD_{N+2}^{(K)}$, respectively, typically will not intersect at a common point. Thus, for finite $K$, checking the geometric inconsistency among $\cC(\btheta_j, \hD_j^{(K)})$, $\cC(\btheta_{N+1}, \hD_{N+1}^{(K)})$ and $\cC(\btheta_{N+2}, \hD_{N+2}^{(K)})$ cannot reliably tell whether the $j$-th sensor is unattacked or not. 
To overcome this, we replace $\cC(\btheta_{N+1}, \hD_{N+1}^{(K)})$ and $\cC(\btheta_{N+2}, \hD_{N+2}^{(K)})$ with  $\cR(\btheta_{N+1}, \hD_{N+1}^{(K)}, \delta)$  and $\cR(\btheta_{N+2}, \hD_{N+2}^{(K)}, \delta)$ for some $\delta$, respectively, and scrutinize whether  $\cC(\btheta_j, \hD_j^{(K)})$ pass through the common area of $\cR(\btheta_{N+1}, \hD_{N+1}^{(K)}, \delta)$  and $\cR(\btheta_{N+2}, \hD_{N+2}^{(K)}, \delta)$ instead.


To be specific, for the $j$-th sensor, $  j=1,2,...,N$, we consider the following hypothesis testing problem
\begin{equation} \label{hypthesis}
\left\{ \begin{array}{l}
{{\cal{H}}_0}: j \in \cal U\\
{{\cal{H}}_1}: j \in \cal V \\
\end{array} \right.
\end{equation}
and a class of detectors
\begin{equation} \label{decision_rule}
\varpi_j \left( \delta \right) = \left\{ \begin{split}
0, & \quad \text{ if } {\cC}\lp \btheta_j, \hD_{j}^{(K)}  \rp \cap \left[ {\mathop  \cap \limits_{i =1}^{2} {\cal R}\left( {{\btheta _{N+i}},\hD_{N+i}^{(K)}}, \delta \right)} \right]   \ne \emptyset,  \\
1, & \quad \text{ if } {\cC}\lp \btheta_j, \hD_{j}^{(K)}  \rp \cap \left[ {\mathop  \cap \limits_{i =1}^{2} {\cal R}\left( {{\btheta _{N+i}},\hD_{N+i}^{(K)}}, \delta \right)} \right]   = \emptyset,
\end{split} \right.
\end{equation}
for some constant $\delta$, where $\hD_{j}^{(K)}$ is defined in (\ref{Definition_hD_jK}). 

\begin{figure}
	\centerline{
		\includegraphics[height=0.5\textwidth]{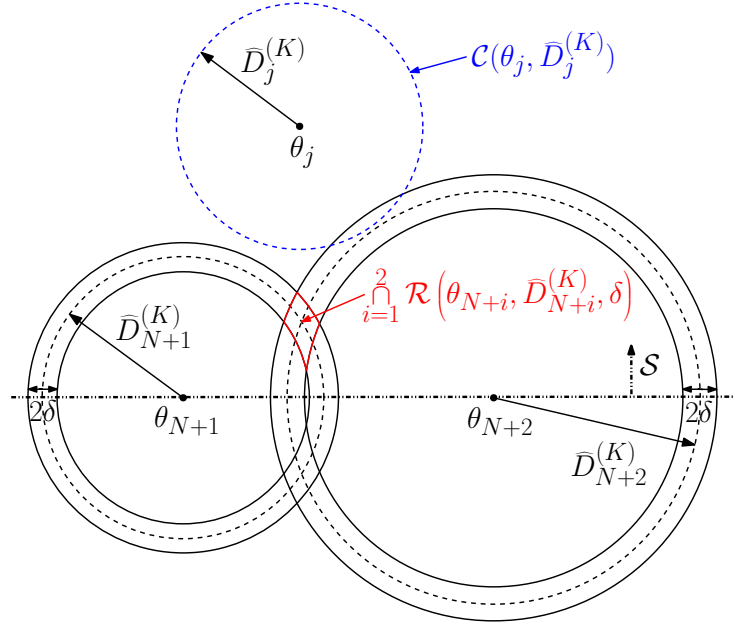}
	}
	\caption{Geometric illustration of the proposed detectors.
	}	
	\label{Fig_Detector}	
\end{figure}

The geometric illustration of the proposed detector in (\ref{decision_rule}) is depicted in Fig. \ref{Fig_Detector}, where the region enclosed by the red curves is the common area of $\cR(\btheta_{N+1}, \hD_{N+1}^{(K)}, \delta)$  and $\cR(\btheta_{N+2}, \hD_{N+2}^{(K)}, \delta)$ which plays an important role in the attack detection process. It is worth noticing that the center of this common area is determined by two random variables $\hD_{N+1}^{(K)}$ and $\hD_{N+2}^{(K)}$, and thereby is randomly located. To this end, this common area may not cover the true target location $\btheta_{\teT}$. In addition, the size of the common area of $\cR(\btheta_{N+1}, \hD_{N+1}^{(K)}, \delta)$  and $\cR(\btheta_{N+2}, \hD_{N+2}^{(K)}, \delta)$ depends on the parameter $\delta$ which impacts the false alarm and miss probabilities of the proposed detector.

\section{Performance Analysis of the Proposed Detector}
\label{Section_Performance_Analysis}

In this section, the detection performance of the proposed detector in (\ref{decision_rule}) is investigated. We will show that the false alarm and miss probabilities of the proposed detector  decay exponentially fast as the number of data samples at each sensor increases.

To start with, we provide the following lemma regarding the lower and upper bounds on the common area of $\cR(\btheta_{N+1}, D_{N+1}, \delta)$ and $\cR(\btheta_{N+2}, D_{N+2}, \delta)$.

\begin{lemma} \label{Lemma_overlap_area}
If 
	\begin{equation}
	\delta < \Upsilon \triangleq \min\{\Upsilon_1, \Upsilon_2 \},
	\end{equation} 
	then
	\begin{equation} \label{Definition_Phi}
	\mathop {\sup }\limits_{\btheta  \in {\mathop  \cap \limits_{i =1}^{2} {\cal R}\left( {{\btheta _{N+i}},\hD_{N+i}^{(K)}}, \delta \right)}  } \left\| {\btheta  - {\btheta _\teT}} \right\| < \Phi(\delta) \triangleq \left( {2{D_U} + \Upsilon } \right)^{\frac{1}{2}}\left[ {\frac{{2{D_U} + \Upsilon }}{{{D_S}}}\left( {\frac{\Upsilon }{{{D_S}}} + 1} \right) + 2} \right]^{\frac{1}{2}}\sqrt{\delta},
	\end{equation}
	which implies 
	\begin{equation}
	\cB(\btheta_\teT, \delta) \subseteq {\mathop  \cap \limits_{i =1}^{2} {\cal R}\left( {{\btheta _{N+i}},\hD_{N+i}^{(K)}}, \delta \right)}  \subseteq \cB \left( \btheta_\teT, \Phi(\delta) \right).
	\end{equation}
\end{lemma}
\begin{IEEEproof}
Refer to Appendix \ref{Proof_Lemma_overlap_area}.
\end{IEEEproof}

As demonstrated by {Lemma \ref{Lemma_overlap_area}}, the common area of  $\cR(\btheta_{N+1}, D_{N+1}, \delta)$ and $\cR(\btheta_{N+2}, D_{N+2}, \delta)$ can be bounded by two balls from  below and above. Moreover, the radii of these two balls are both increasing functions of the given $\delta$. It will be shown later that by employing the two balls to approximate the irregular area ${ \cap_{i =1}^{2} {\cal R}( {{\btheta _{N+i}},\hD_{N+i}^{(K)}}, \delta )}$ from below and above, the detection performance analysis of the proposed detector in (\ref{decision_rule}) can be considerably facilitated. 

\subsection{Upper Bound on False Alarm Probability}

From (\ref{decision_rule}), the false alarm and miss probabilities of the proposed detector are given by ${\bbP_0}\left( {{\varpi_j \left( \delta \right) =1}} \right)$ and ${\bbP_1}\left( {{\varpi_j \left( \delta \right) =0}} \right)$, respectively, where $\bbP_i$ denotes the probability measure under hypothesis ${\mathcal H}_i$.

Let $\cE_i$ denote the event
\begin{equation} \label{Definition_E_i}
{\cal E}_i \buildrel \Delta \over = \left\{ {\left| {\hD_{N+i}^{(K)} - D_{N+i}} \right| < \frac{1}{2}\delta} \right\}, \;   i = 1,2,
\end{equation}
and ${\cE_i^{\rm{C}}}$ denotes the complement of the event $\cE_i$. The false alarm probability of the detector in (\ref{decision_rule}) can be expressed as
\begin{align} \notag
{\bbP_0}\left( {{\varpi_j \left( \delta \right) =1}} \right) & = {\bbP_0}\left( {   {\cal C}\left( {{\btheta _j},\hD_j^{(K)}} \right) \cap \left[ {\mathop  \cap \limits_{i =1}^{2} {\cal R}\left( {{\btheta _{N+i}},\hD_{N+i}^{(K)}}, \delta \right)} \right] = \emptyset  }  \right)\\ \notag
& = {\bbP_0}\left( { \lcb {\cal C}\left( {{\btheta _j},\hD_j^{(K)}} \right) \cap \left[ {\mathop  \cap \limits_{i = 1}^{2} {\cal R}\left( {{\btheta _{N+i}},\hD_{N+i}^{(K)}}, \delta \right)} \right] = \emptyset \rcb \cap \lp \cE_1 \cap \cE_2  \rp  }  \right)\\ \label{Pfa_1}
& \quad + {\bbP_0}\left(  { \lcb  {\cal C}\left( {{\btheta _j},\hD_j^{(K)}} \right) \cap \left[ {\mathop  \cap \limits_{i =1}^{2} {\cal R}\left( {{\btheta _{N+i}},\hD_{N+i}^{(K)}}, \delta \right)} \right]  = \emptyset \rcb \cap \lp {\cE_1^{\rm{C}}}\cup {\cE_2^{\rm{C}}} \rp  } \right).
\end{align}

Note that 
${\cal E}_i$ implies that
\begin{equation}
\cR\lp \btheta_{N+i}, D_{N+i}, \frac{1}{2}\delta \rp \subseteq \cR\lp \btheta_{N+i}, \hD_{N+i}^{(K)}, \delta \rp,
\end{equation}
and hence, from (\ref{Pfa_1}), we can obtain
\begin{align} \notag
{\bbP_0}\left( {{\varpi_j \left( \delta \right) =1}} \right)  & \le {\bbP_0}\left( {\left\{ {{\cal C}\left( {{\btheta _j}, \hD_j^{(K)}} \right) \cap \left[ {\mathop  \cap \limits_{i = 1}^2 {\cal R}\left( {{\btheta _{N + i}},{D_{N + i}},\frac{1}{2}\delta } \right)} \right] = \emptyset } \right\} \cap \left( {{{\cal E}_1} \cap {{\cal E}_2}} \right)} \right)\\ \notag
& \quad + {\bbP_0}\left( {\left\{ {{\cal C}\left( {{\btheta _j},\hD_j^{(K)}} \right) \cap \left[ {\mathop  \cap \limits_{i = 1}^2 {\cal R}\left( {{\btheta _{N + i}},\hD_{N + i}^{(K)},\delta } \right)} \right] = \emptyset } \right\} \cap \left( {{\cal E}_1^{\rm{C}} \cup {\cal E}_2^{\rm{C}}} \right)} \right)\\ \label{temp_1}
& \le {\bbP_0}\left( {{\cal C}\left( {{\btheta _j},\hD_j^{(K)}} \right) \cap \left[ {\mathop  \cap \limits_{i = 1}^2 {\cal R}\left( {{\btheta _{N + i}},{D_{N + i}},\frac{1}{2}\delta } \right)} \right] = \emptyset } \right) + {\bbP_0}\left( {{\cal E}_1^{\rm{C}} \cup {\cal E}_2^{\rm{C}}} \right)\\
& \le {\bbP_0}\left( {{\cal C}\left( {{\btheta _j},\hD_j^{(K)}} \right) \cap \left[ {\mathop  \cap \limits_{i = 1}^2 {\cal R}\left( {{\btheta _{N + i}},{D_{N + i}},\frac{1}{2}\delta } \right)} \right]\! =\! \emptyset } \right) \! +\! {\bbP_0}\left( {{\cal E}_1^{\rm{C}}} \right) \! +\! {\bbP_0}\left( {{\cal E}_2^{\rm{C}}} \right),
\end{align}
where (\ref{temp_1}) is due to the fact that ${\bbP}_0({\cal E} \cap {\cal F}) \le {\bbP}_0 ({\cal E})$ for any two events $\cal E$ and $\cal F$.
Moreover, from {Lemma \ref{Lemma_overlap_area}}, we know
\begin{equation}
\cB \lp \btheta_{\rm{T}}, \frac{1}{2}\delta \rp \subseteq {\mathop  \cap \limits_{i = 1}^2 {\cal R}\left( {{\btheta _{N + i}},{D_{N + i}},\frac{1}{2}\delta } \right)},
\end{equation}
which yields
\begin{equation} \label{Pfa_2}
{\bbP_0}\left( {{\varpi_j \left( \delta \right) =1}} \right) \le {\bbP_0}\left( {{\cal C}\left( {{\btheta _j},\hD_j^{(K)}} \right) \cap \cB \lp \btheta_{\rm{T}}, \frac{1}{2}\delta \rp  = \emptyset } \right) + {\bbP_0}\left( {{\cal E}_1^{\rm{C}}} \right) + {\bbP_0}\left( {{\cal E}_2^{\rm{C}}} \right).
\end{equation}
%
%

\begin{figure}
	\centerline{
		\includegraphics[height=0.4\textwidth]{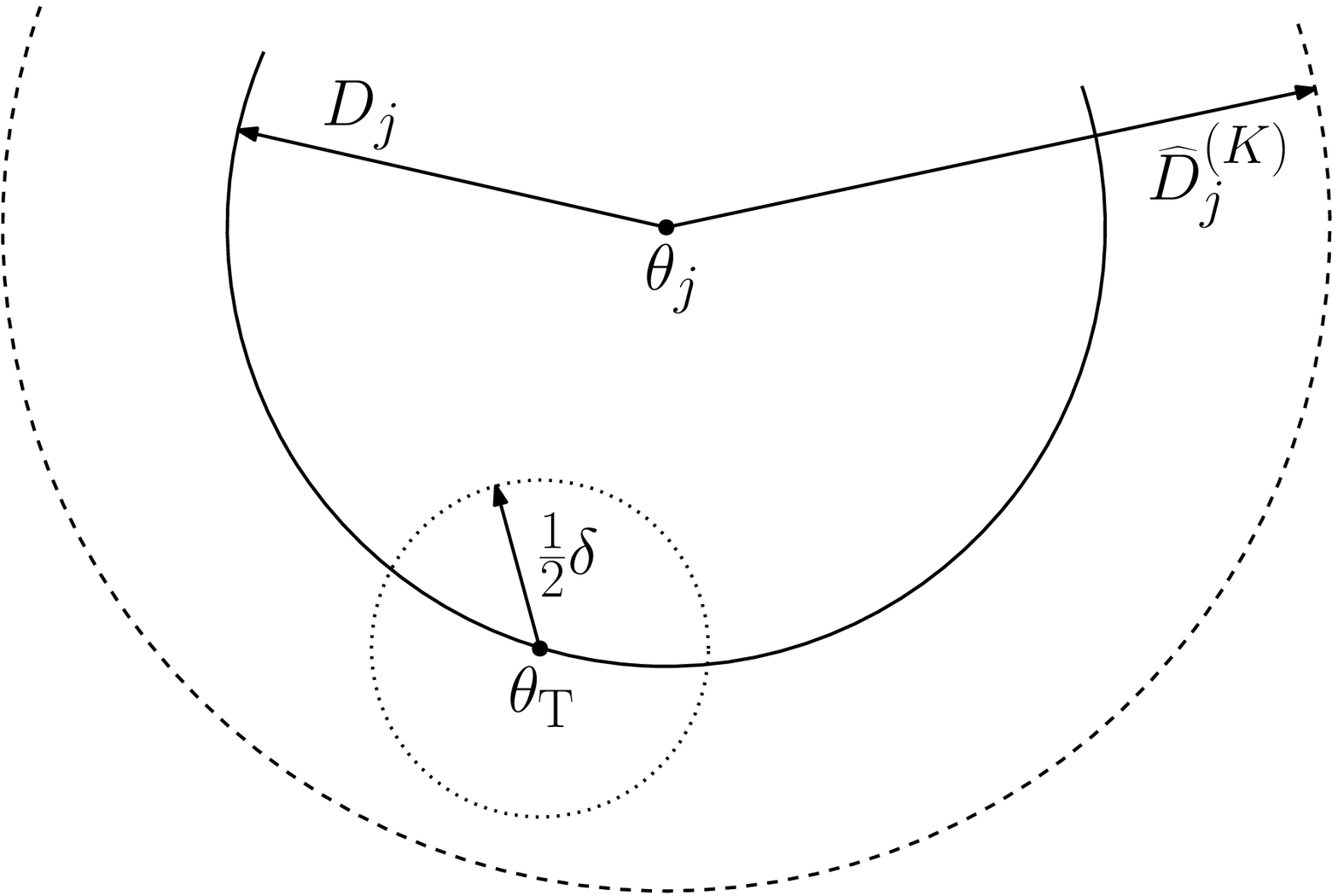}
	}
	\caption{Geometric illustration of (\ref{temp_P_fa_1}).
	}	
	\label{Fig_temp_P_fa_1}	
\end{figure}

In addition, as illustrated in Fig. \ref{Fig_temp_P_fa_1}, if $j \in \cU$, we know $\btheta_{\rm T} \in {\cal C}( {{\btheta _j},D_j} )$ which yields that under hypothesis $\cH_0$,
\begin{equation} \label{temp_P_fa_1}
\left\{ { {\cal C}\left( {{\btheta _j},\hD_j^{(K)}} \right) \cap {\cal B}\left( {{\btheta _{\rm{T}}},\frac{1}{2}\delta } \right) = \emptyset } \right\} \Leftrightarrow \left\{ {\left| {\hD_j^{(K)} - {D_j}} \right| \ge \frac{1}{2}\delta } \right\},
\end{equation}
and therefore, by employing (\ref{Definition_E_i}) and (\ref{Pfa_2}), the false alarm probability can be bounded from above as per
\begin{align} \notag
{\bbP_0}\left( {{\varpi_j \left( \delta \right) =1}} \right) & \le {\bbP_0}\left( {\left| {\hD_j^{(K)} - {D_j}} \right| \ge \frac{1}{2}\delta } \right) + {\bbP_0}\left( {{\cal E}_1^{\rm{C}}} \right) + {\bbP_0}\left( {{\cal E}_2^{\rm{C}}} \right) \\ \label{Pfa_upper_bound}
& = {\bbP_0}\left( { \left| {\hD_j^{(K)} - {D_j}} \right| \ge \frac{1}{2}\delta  } \right) + \sum\limits_{i = 1}^2 {{\bbP_0}\left( {\left| {\hD_{N + i}^{(K)} - {D_{N + i}}} \right| \ge \frac{1}{2}\delta } \right)}.
\end{align}

\subsection{Upper Bound on Miss Probability}

On the other hand, the miss probability of the detector in (\ref{decision_rule}) can be bounded from above as per
\begin{align} \notag
{\bbP_1}\left( {{\varpi_j \left( \delta \right) =0}} \right) & = {\bbP_1}\left( {{\cal C}\left( {{\btheta _j},\hD_j^{(K)}} \right) \cap \left[ {\mathop  \cap \limits_{i = 1}^2 {\cal R}\left( {{\btheta _{N + i}},\hD_{N + i}^{(K)},\delta } \right)} \right] \ne \emptyset } \right)\\ \notag
& = {\bbP_1}\left( {\left\{ {{\cal C}\left( {{\btheta _j},\hD_j^{(K)}} \right) \cap \left[ {\mathop  \cap \limits_{i = 1}^2 {\cal R}\left( {{\btheta _{N + i}},\hD_{N + i}^{(K)},\delta } \right)} \right] \ne \emptyset } \right\} \cap \left( {{{\cal E}_1} \cap {{\cal E}_2}} \right)} \right)\\ \notag
& \quad + {\bbP_1}\left( {\left\{ {{\cal C}\left( {{\btheta _j},\hD_j^{(K)}} \right) \cap \left[ {\mathop  \cap \limits_{i = 1}^2 {\cal R}\left( {{\btheta _{N + i}},\hD_{N + i}^{(K)},\delta } \right)} \right] \ne \emptyset } \right\} \cap \left( {{\cal E}_1^{\rm{C}} \cup {\cal E}_2^{\rm{C}}} \right)} \right) \\ \notag
& \le {\bbP_1}\left( {\left\{ {{\cal C}\left( {{\btheta _j},\hD_j^{(K)}} \right) \cap \left[ {\mathop  \cap \limits_{i = 1}^2 {\cal R}\left( {{\btheta _{N + i}},{D_{N + i}},\frac{3}{2}\delta } \right)} \right] \ne \emptyset } \right\} \cap \left( {{{\cal E}_1} \cap {{\cal E}_2}} \right)} \right)\\ \label{Pmd_temp1}
&\quad  + {\bbP_1}\left( {{\cal E}_1^{\rm{C}} \cup {\cal E}_2^{\rm{C}}} \right)\\ \notag
& \le {\bbP_1}\left( {{\cal C}\left( {{\btheta _j},\hD_j^{(K)}} \right) \cap \left[ {\mathop  \cap \limits_{i = 1}^2 {\cal R}\left( {{\btheta _{N + i}},{D_{N + i}},\frac{3}{2}\delta } \right)} \right] \ne \emptyset } \right)  + {\bbP_1}\left( {{\cal E}_1^{\rm{C}}} \right) + {\bbP_1}\left( {{\cal E}_2^{\rm{C}}} \right)\\ \label{Pmd_temp2}
& \le {\bbP_1}\left( {{\cal C}\left( {{\btheta _j},\hD_j^{(K)}} \right) \cap {\cal B}\left( {{\btheta _{\rm{T}}},\Phi \left( {\frac{3}{2}\delta } \right)} \right) \ne \emptyset } \right) + {\bbP_1}\left( {{\cal E}_1^{\rm{C}}} \right) + {\bbP_1}\left( {{\cal E}_2^{\rm{C}}} \right),
\end{align}
where (\ref{Pmd_temp1}) is due to the fact that if $\cE_1$ and $\cE_2$ occur, then
\begin{equation}
{\cal R}\left( {{\btheta _{N + i}},\hD_{N + i}^{(K)},\delta } \right) \subseteq {\cal R}\left( {{\btheta _{N + i}},D_{N + i}, \frac{3}{2}\delta } \right), \;   i=1,2, 
\end{equation}
and (\ref{Pmd_temp2}) is because ${\cap_{i = 1}^2 {\cal R}\left( {{\btheta _{N + i}},{D_{N + i}},\frac{3}{2}\delta } \right)}  \subseteq {\cal B}\left( {{\btheta _{\rm{T}}},\Phi \left( {\frac{3}{2}\delta } \right)} \right)$ according to {Lemma \ref{Lemma_overlap_area}}.

Since the first term in (\ref{Pmd_temp2}) is hard to deal with, we employ an upper bound on it which is provided in the following lemma.

\begin{lemma} \label{Lemma_Pmd_part1}
	Define
	\begin{equation} \label{Definition_lambda_j}
	 \lambda_j \triangleq \frac{\kappa {{D_0}P_0^{\frac{1}{\gamma }}{{\left[ {{\tau _j} - F_j^{ - 1}\left( {{\rho_{j}^{(\rm{L})}}} \right)} \right]}^{ - \frac{{\gamma  + 1}}{\gamma }}}}}{{\mathop {\sup }\limits_{x \in \left[ {{F_j^{-1}(\rho_{j}^{(\rm{L})})},{F_j^{-1}(\rho_{j}^{(\rm{U})})}} \right]} {f_j}\left( x \right)}}, 
	\end{equation}
	and denote 
	\begin{equation} \label{Definition_lambda}
	\lambda  = \mathop {\min }\limits_{j = 1,2,...,N} \left\{ {{\lambda _j}} \right\}.
	\end{equation}
If 
	\begin{equation} \label{lemma_delta_constraint}
	0< \delta  < \min {\left\{ {\Upsilon ,\left\{ {{{\left( {2{D_U} + \Upsilon } \right)}^{\frac{1}{2}}}{{\left[ {\frac{{6{D_U} + 3\Upsilon }}{{2{D_S}}}\left( {\frac{\Upsilon }{{{D_S}}} + 1} \right) + 3} \right]}^{\frac{1}{2}}} + {\frac{1}{2}\Upsilon ^{\frac{1}{2}}}} \right\}^{ - 2}{\lambda ^2}} \right\}},
	\end{equation}
	then
	\begin{equation}
	{\bbP_1}\left( {{\cal C}\left( {{\btheta _j},\hD_j^{(K)}} \right) \cap {\cal B}\left( {{\btheta _{\rm{T}}},\Phi \left( {\frac{3}{2}\delta } \right)} \right) \ne \emptyset } \right)  \le {\bbP_1}\left( {\left| {\hD_j^{(K)} - {{\widetilde D}_j}} \right| \ge  \frac{1}{2}\delta } \right),
	\end{equation}
	where ${{\widetilde D}_j}$ is defined in (\ref{Define_tilde_D_j}).
\end{lemma}

\begin{IEEEproof}
Refer to Appendix \ref{Proof_Lemma_Pmd_part1}.
\end{IEEEproof}

It is worth mentioning that since $f_j(x)$ is continuous and positive over $\{x | F_j(x) \in (0,1) \}$,  the denominator  ${ {\sup }_{x \in \left[ {{F_j^{-1}(\rho_{j}^{(\rm{L})})},{F_j^{-1}(\rho_{j}^{(\rm{U})})}} \right]} {f_j}\left( x \right)}$ in (\ref{Definition_lambda_j}) is positive and bounded. Moreover, according to (\ref{Definition_rho_L}), we know that $\tau_j > F_j^{-1}(\rho_{j}^{(\rm{L})})$, since $F_j^{-1}$ is strictly increasing. Therefore, $0< \lambda_j < \infty$, and hence, $0< \lambda< \infty$.

By employing (\ref{Pmd_temp2}) and Lemma \ref{Lemma_Pmd_part1}, we know if (\ref{lemma_delta_constraint}) holds, 
then an upper bound on the miss probability of the detector in (\ref{decision_rule}) can be expressed as
\begin{align} \notag
{\bbP_1}\left( {{\varpi_j \left( \delta \right) =0}} \right)  & \le  {\bbP_1}\left( {\left| {\hD_j^{(K)} - {{\widetilde D}_j}} \right| \ge  \frac{1}{2}\delta } \right) + {\bbP_1}\left( {{\cal E}_1^{\rm{C}}} \right) + {\bbP_1}\left( {{\cal E}_2^{\rm{C}}} \right) \\ \label{Pmd_upper_bound}
& = {\bbP_1}\left( {\left| {\hD_j^{(K)} - {{\widetilde D}_j}} \right| \ge  \frac{1}{2}\delta } \right)  + \sum\limits_{i = 1}^2 {{\bbP_1}\left( {\left| {\hD_{N + i}^{(K)} - {D_{N + i}}} \right| \ge \frac{1}{2}\delta } \right)}.
\end{align}

\subsection{Exponential Decay of False Alarm and Miss Probabilities}

It is seen from (\ref{Pfa_upper_bound}) and (\ref{Pmd_upper_bound}) that 
the upper bounds on the false alarm and miss probabilities have some similarities. To be specific, since the $(N+1)$-th and $(N+2)$-th sensors are secure, ${{\bbP_0}\left( {\left| {\hD_{N + i}^{(K)} - {D_{N + i}}} \right| \ge \frac{1}{2}\delta } \right)}={{\bbP_1}\left( {\left| {\hD_{N + i}^{(K)} - {D_{N + i}}} \right| \ge \frac{1}{2}\delta } \right)}$ for $i=1,2$. Thus, the second term in (\ref{Pfa_upper_bound}) is the same as the second term in  (\ref{Pmd_upper_bound}). Moreover, as $K \to \infty$,   $\hD_j^{(K)} \to D_j$ almost surely under hypothesis $\cH_0$, while $\hD_j^{(K)} \to \tD_j$ almost surely under hypothesis $\cH_1$,  one can expect that the first term in (\ref{Pfa_upper_bound}) and the first term in (\ref{Pmd_upper_bound}) behave in a very similar way as $K$ increases, except for the change in $\hD_j^{(K)}$ due to the attack. In the following theorem, by employing (\ref{Pfa_upper_bound}) and (\ref{Pmd_upper_bound}), we show that the false alarm and miss probabilities of the detector in (\ref{decision_rule}) decay at least  exponentially with respect to $K$.

\begin{theorem} \label{Theorem_Performance}
	If (\ref{lemma_delta_constraint}) holds,
	then the false alarm and miss probabilities are upper bounded by
	\begin{equation}
	{\bbP_0}\left( {{\varpi_j \left( \delta \right) =1}} \right) \le 12{e^{ - \eta _j^{({0})}\left( \delta  \right)K}},
	\end{equation}
	\begin{equation}
	{\bbP_1}\left( {{\varpi_j \left( \delta \right) =0}} \right) \le 12{e^{ - \eta _j^{({1})}\left( \delta  \right)K}},
	\end{equation}
	for some positive constants $\eta _j^{(0)}\left( \delta \right)$ and $\eta _j^{(1)}\left( \delta \right)$.
\end{theorem}

\begin{IEEEproof}
Before proceeding, we define a sequence of events $\cF_{j,K}$ as
\begin{equation} \label{Define_F_jk}
\cF_{j,K}  \triangleq \lcb  \xi_j^{(K)} \in \lsb {\varepsilon _j^{({\rm{L}})}},  {\varepsilon _j^{({\rm{U}})}} \rsb  \rcb,
\end{equation}
where  $\xi_j^{(K)}$ is defined in (\ref{Definition_xi_jK}).  The constants ${\varepsilon _j^{({\rm{L}})}}$ and ${\varepsilon _j^{({\rm{U}})}}$ in (\ref{Define_F_jk}) are defined as
\begin{equation} \label{Definition_epsilon_j}
\varepsilon _j^{({\rm{L}})} \buildrel \Delta \over = \sigma _j^{({\rm{L}})}{\rho_{j}^{(\rm{L})}} \quad \text{and} \quad \varepsilon _j^{({\rm{U}})} \buildrel \Delta \over = \sigma _j^{({\rm{U}})}{\rho_{j}^{(\rm{U})}} + \left( {1 - \sigma _j^{({\rm{U}})}} \right){F_j}\left( {{\tau _j}} \right)
\end{equation}
for some numbers $\sigma _j^{({\rm{L}})}, \sigma _j^{({\rm{U}})} \in (0,1)$, where $F_j\left( \tau_j \right){\rm{ }} \buildrel \Delta \over = \int_{ - \infty }^{\tau_j} {f_j\left( t \right)dt} $, and ${\rho_{j}^{(\rm{L})}}$ and ${\rho_{j}^{(\rm{U})}}$ are defined in (\ref{Definition_rho_L}) and (\ref{Definition_rho_U}), respectively. 
From (\ref{rho_L_rho_U_less_1}) and (\ref{Definition_epsilon_j}),
we know that
\begin{equation} \label{epsilon_range}
0 < {\varepsilon _j^{({\rm{L}})}}  <  \rho_{j}^{(\rm{L})} < \rho_{j}^{(\rm{U})} < {\varepsilon _j^{({\rm{U}})}} < F_j\lp \tau_j \rp \le 1.
\end{equation}
	
Let's first consider the upper bound on the false alarm probability as illustrated in (\ref{Pfa_upper_bound}). For any $j \in \{1,2,...,N+2\}$,
\begin{align} \notag
{\bbP_0}\left( {\left| {\hD_j^{(K)} - {D_j}} \right| \ge \frac{1}{2}\delta } \right) & = {\bbP_0}\left( {\left\{ {\left| {\hD_j^{(K)} - {D_j}} \right| \ge \frac{1}{2}\delta } \right\} \cap {{\cal F}_{j,K}}} \right)\\ \notag
& \quad  + {\bbP_0}\left( {\left\{ {\left| {\hD_j^{(K)} - {D_j}} \right| \ge \frac{1}{2}\delta } \right\} \cap {\cal F}_{j,K}^{\rm{C}}} \right) \\ \label{P_0_D_j_K_D_j}
& \le {\bbP_0}\left( {\left\{ {\left| {\hD_j^{(K)} - {D_j}} \right| \ge \frac{1}{2}\delta } \right\} \cap {{\cal F}_{j,K}}} \right)  + {\bbP_0}\left( {\cal F}_{j,K}^{\rm{C}} \right).
\end{align}

Note that under hypothesis $\cH_0$, if $\xi_j^{(K)} \in [  {\varepsilon _j^{({\rm{L}})}},  {\varepsilon _j^{({\rm{U}})}} ]$, then by employing (\ref{D_j}), (\ref{Definition_hD_jK}) and (\ref{epsilon_range}), we can obtain
\begin{align} \notag
\left| {\hD_j^{(K)} - {D_j}} \right| & = {D_0}P_0^{\frac{1}{\gamma }}\left| {{{\left[ {{\tau _j} - F_j^{ - 1}\left( {\xi _j^{(K)}} \right)} \right]}^{ - \frac{1}{\gamma }}} - {{\left[ {{\tau _j} - F_j^{ - 1}\left( {{p_j}\left( {{\btheta _{\rm{T}}}} \right)} \right)} \right]}^{ - \frac{1}{\gamma }}}} \right|\\ \label{temp_2}
& \le {D_0}P_0^{\frac{1}{\gamma }}\mathop {\sup }\limits_{x \in \left[ {{\varepsilon _j^{({\rm{L}})}},{\varepsilon _j^{({\rm{U}})}}} \right]} \left| {\frac{{\partial {{\left[ {{\tau _j} - F_j^{ - 1}\left( x \right)} \right]}^{ - \frac{1}{\gamma }}}}}{{\partial x}}} \right|\left| {\xi _j^{(K)} - {p_j}\left( {{\btheta _{\rm{T}}}} \right)} \right|\\ \notag
& = {D_0}P_0^{\frac{1}{\gamma }}\mathop {\sup }\limits_{x \in \left[ {{\varepsilon _j^{({\rm{L}})}},{\varepsilon _j^{({\rm{U}})}}} \right]} \left| {\frac{{{{\left[ {{\tau _j} - F_j^{ - 1}\left( x \right)} \right]}^{ - \frac{{\gamma  + 1}}{\gamma }}}}}{{{f_j}\left( {F_j^{ - 1}\left( x \right)} \right)}}} \right|\left| {\xi _j^{(K)} - {p_j}\left( {{\btheta _{\rm{T}}}} \right)} \right|\\  \label{D_jk_D_j_diff}
& \le   \underbrace {\frac{{{D_0}P_0^{\frac{1}{\gamma }}{{\left[ {{\tau _j} - F_j^{ - 1}\left( {\varepsilon _j^{({\rm{U}})}} \right)} \right]}^{ - \frac{{\gamma  + 1}}{\gamma }}}}}{{\mathop {\inf }\limits_{x \in \left[ {F_j^{ - 1}\left( {\varepsilon _j^{({\rm{L}})}} \right), F_j^{ - 1}\left( {\varepsilon _j^{({\rm{U}})}} \right)} \right]} {f_j}\left( x \right)}}}_{{\Xi _j}}\left| {\xi _j^{(K)} - {p_j}\left( {{\btheta _{\rm{T}}}} \right)} \right|,
\end{align}
where (\ref{temp_2}) is due to the fact that $p_j(\btheta_{\teT}) \in  [  {\varepsilon _j^{({\rm{L}})}},  {\varepsilon _j^{({\rm{U}})}} ]$ and $\xi_j^{(K)} \in [  {\varepsilon _j^{({\rm{L}})}},  {\varepsilon _j^{({\rm{U}})}} ]$.
%
%
Since $f_j(x)$ is continuous and $0<f_j(x)<\infty$ over $\{x | F_j(x) \in (0,1) \}$, we know
\begin{equation}
{\mathop {\inf }\limits_{x \in \left[ {F_j^{ - 1}\left( {\varepsilon _j^{({\rm{L}})}} \right), F_j^{ - 1}\left( {\varepsilon _j^{({\rm{U}})}} \right)} \right]} {f_j}\left( x \right)} =  {\mathop {\min }\limits_{x \in \left[ {F_j^{ - 1}\left( {\varepsilon _j^{({\rm{L}})}} \right), F_j^{ - 1}\left( {\varepsilon _j^{({\rm{U}})}} \right)} \right]} {f_j}\left( x \right)} \in (0, \infty),
\end{equation}
and moreover, from (\ref{epsilon_range}), we know
\begin{equation}
{{\tau _j} - F_j^{ - 1}\left( {\varepsilon _j^{({\rm{U}})}} \right)} > 0,
\end{equation}
since ${\varepsilon _j^{({\rm{U}})}}  < F_j\lp \tau_j \rp$. Therefore, it is clear that $\Xi_j \in (0, \infty)$.

By employing (\ref{D_jk_D_j_diff}), we can obtain
\begin{align} \notag
&{\bbP_0}\left( {\left\{ {\left| {D_j^{(K)} - {D_j}} \right| \ge \frac{1}{2}\delta } \right\} \cap {{\cal F}_{j,K}}} \right) \\ \notag
& \le {\bbP_0}\left( {\left\{ {{\Xi _j}\left| {\xi _j^{(K)} - {p_j}\left( {{\btheta _{\rm{T}}}} \right)} \right| \ge \frac{1}{2}\delta } \right\} \cap {{\cal F}_{j,K}}} \right)\\ \notag
&\le {\bbP_0}\left( {\xi _j^{(K)} - {p_j}\left( {{\btheta _{\rm{T}}}} \right) \ge \frac{\delta }{{2{\Xi _j}}}} \right) + {\bbP_0}\left( {\xi _j^{(K)} - {p_j}\left( {{\btheta _{\rm{T}}}} \right) \le  - \frac{\delta }{{2{\Xi _j}}}} \right)\\ \label{P_0_Dkj_D_j}
&= {\bbP_0}\left( {\sum\limits_{k = 1}^K {\underbrace { \lp  1 - {{\tilde u}_{jk}} - {p_j}\left( {{\btheta _{\rm{T}}}} \right) \rp }_{{X_{jk}}}}  \ge \frac{\delta }{{2{\Xi _j}}}K} \right) + {\bbP_0}\left( {\sum\limits_{k = 1}^K {\underbrace { \lp  {{\tilde u}_{jk}} + {p_j}\left( {{\btheta _{\rm{T}}}} \right) -1\rp }_{{Y_{jk}}} }  \ge   \frac{\delta }{{2{\Xi _j}}}K} \right).
\end{align}
It is easy to see that under hypothesis $\cH_0$, $\{X_{jk}\}_{k=1}^K$  is a sequence of independent and identically distributed random variables with distribution
\begin{align} \label{Definition_q_Xjk}
{q_{{X_{j}}}} & \buildrel \Delta \over = {\bbP_0}\left( {{X_{jk}} = 1 - {p_j}\left( {{\btheta _{\rm{T}}}} \right)} \right) = {p_j}\left( {{\btheta _{\rm{T}}}} \right)\\ \label{Definition_q_Xjk_bar}
{{\bar q}_{{X_{j}}}} & \buildrel \Delta \over = {\bbP_0}\left( {{X_{jk}} = - {p_j}\left( {{\btheta _{\rm{T}}}} \right)} \right) = 1 - {p_j}\left( {{\btheta _{\rm{T}}}} \right).
\end{align}
Since
\begin{equation}
\frac{\delta }{{2{\Xi _j}}}K > {\bbE_0}\left\{ {{X_{jk}}} \right\} = \lsb 1 - {p_j}\left( {{\btheta _{\rm{T}}}} \right) \rsb  {q_{{X_{j}}}} - {p_j}\left( {{\btheta _{\rm{T}}}} \right) {{\bar q}_{{X_{j}}}} = 0,
\end{equation}
by employing the large deviations theory \cite{dembo2010large}, we can obtain 
\begin{equation} \label{P_0_Dkj_D_j_1}
{\bbP_0}\left( {\sum\limits_{k = 1}^K { \lp 1 - {{\tilde u}_{jk}} - {p_j}\left( {{\btheta _{\rm{T}}}} \right) \rp }  \ge {\frac{\delta }{{2{\Xi _j}}}}K} \right) \le {e^{ - {  \eta _{j,1}\left( \delta  \right)  }K}},
\end{equation}
where the rate function $\eta _{j,1}\left( \delta  \right)$ is defined as
\begin{align} \notag
\eta _{j,1}\left( \delta  \right) & \buildrel \Delta \over =  - \mathop {\lim }\limits_{K \to \infty } \frac{1}{K}\ln {\bbP_0}\left( {\sum\limits_{k = 1}^K {{X_{jk}}}  \ge {\frac{\delta }{{2{\Xi _j}}}}K} \right) \\ \label{Definition_eta_j}
& = {\frac{\delta }{{2{\Xi _j}}}}{\mu ^*} - \ln \phi_{X_j} \left( {{\mu ^*}} \right),
\end{align}
\begin{equation}
\text{and} \quad {\phi _{{X_j}}}\left( \mu  \right) \buildrel \Delta \over =  {\bbE_0}\left\{ {{e^{\mu {X_{jk}}}}} \right\} =  {{p_j}\left( {{\btheta _{\rm{T}}}} \right){e^{\mu \left( {1 - {p_j}\left( {{\btheta _{\rm{T}}}} \right)} \right)}} + \left( {1 - {p_j}\left( {{\btheta _{\rm{T}}}} \right)} \right){e^{ - \mu {p_j}\left( {{\btheta _{\rm{T}}}} \right)}}} .
\end{equation}
Moreover, the quantity $\mu^*$ in (\ref{Definition_eta_j}) is the solution of the equation 
\begin{equation} \label{equation_eta_star}
\frac{d}{{d\mu }}{\phi _{{X_j}}}\left( \mu  \right) = { \frac{\delta }{{2{\Xi _j}}}  }{\phi _{{X_j}}}\left( \mu  \right).
\end{equation}
By employing (\ref{Definition_eta_j})--(\ref{equation_eta_star}), the rate function $\eta _{j,1}\left( \delta  \right)$ can be obtained as
\begin{equation}  \label{eta_j_result}
\eta _{j,1}\left( \delta  \right) =  \eta _{j,1}^*\left( \delta  \right) \triangleq \left( {\frac{\delta }{{2{\Xi _j}}} + {p_j}\left( {{\btheta _{\rm{T}}}} \right)} \right)\ln \frac{{\left( {\frac{\delta }{{2{\Xi _j}}} + {p_j}\left( {{\btheta _{\rm{T}}}} \right)} \right)\left( {1 - {p_j}\left( {{\btheta _{\rm{T}}}} \right)} \right)}}{{{p_j}\left( {{\btheta _{\rm{T}}}} \right)\left( {1 - \frac{\delta }{{2{\Xi _j}}} - {p_j}\left( {{\btheta _{\rm{T}}}} \right)} \right)}} - \ln \frac{{1 - {p_j}\left( {{\btheta _{\rm{T}}}} \right)}}{{1 - \frac{\delta }{{2{\Xi _j}}} - {p_j}\left( {{\btheta _{\rm{T}}}} \right)}},
\end{equation}
provided that 
\begin{equation}
\frac{\delta }{{2{\Xi _j}}} \le {1  - {p_j}\left( {{\btheta _{\rm{T}}}} \right)}.
\end{equation}
It is seen from (\ref{Definition_q_Xjk}) and (\ref{Definition_q_Xjk_bar}) that
\begin{equation}
\sum\limits_{k = 1}^K { \lp 1 - {{\tilde u}_{jk}} - {p_j}\left( {{\btheta _{\rm{T}}}} \right) \rp } \le \lp 1  - {p_j}\left( {{\btheta _{\rm{T}}}} \right) \rp K,
\end{equation}
which implies that for the case where $\frac{\delta }{{2{\Xi _j}}} > {1  - {p_j}\left( {{\btheta _{\rm{T}}}} \right)}$,
\begin{equation}
{\bbP_0}\left( {\sum\limits_{k = 1}^K { \lp 1 - {{\tilde u}_{jk}} - {p_j}\left( {{\btheta _{\rm{T}}}} \right) \rp }  \ge {\frac{\delta }{{2{\Xi _j}}}}K} \right) =0.
\end{equation}
Therefore, the rate function $\eta _{j,1}\left( \delta  \right)$ can be written as\footnote{Regarding the second term of the right-hand side of  (\ref{eta_j_1_0}), we define $\infty\cdot0=0$.}
\begin{equation} \label{eta_j_1_0}
\eta _{j,1}\left( \delta  \right) =  \eta _{j,1}^*\left( \delta  \right)\bone\left\{ {\frac{\delta }{{2{\Xi _j}}} \le 1 - {p_j}\left( {{\btheta _{\rm{T}}}} \right)} \right\} + \infty \bone\left\{ {\frac{\delta }{{2{\Xi _j}}} > 1 - {p_j}\left( {{\btheta _{\rm{T}}}} \right)} \right\},
\end{equation}
where $ \eta _{j,1}^*\left( \delta  \right)$ is defined in (\ref{eta_j_result}).

Similarly, noting that under hypothesis $\cH_0$, $\{Y_{jk}\}_{k=1}^K$  is a sequence of independent and identically distributed random variables with distribution
\begin{align} \label{Definition_q_Yjk}
{q_{{X_{j}}}} & \buildrel \Delta \over = {\bbP_0}\left( {{Y_{jk}} =  {p_j}\left( {{\btheta _{\rm{T}}}} \right) -1 } \right) = {p_j}\left( {{\btheta _{\rm{T}}}} \right)\\ \label{Definition_q_Yjk_bar}
{{\bar q}_{{X_{j}}}} & \buildrel \Delta \over = {\bbP_0}\left( {{Y_{jk}} = {p_j}\left( {{\btheta _{\rm{T}}}} \right)} \right) = 1 - {p_j}\left( {{\btheta _{\rm{T}}}} \right),
\end{align}
we can obtain 
\begin{equation} \label{P_0_Dkj_D_j_2}
{\bbP_0}\left( {\sum\limits_{k = 1}^K { \lp {{\tilde u}_{jk}} + {p_j}\left( {{\btheta _{\rm{T}}}} \right) -1 \rp }  \ge {\frac{\delta }{{2{\Xi _j}}}}K} \right) \le {e^{ - {  \eta _{j,2}\left( \delta  \right)  }K}},
\end{equation}
where the rate function $\eta _{j,2}\left( \delta  \right)$ is given by
\begin{equation} \label{eta_j_2_0}
\eta _{j,2}\left( \delta  \right) =  \eta _{j,2}^*\left( \delta  \right){\bone}\left\{ {\frac{\delta }{{2{\Xi _j}}} \le {p_j}\left( {{\btheta _{\rm{T}}}} \right)} \right\} + \infty {\bone}\left\{ {\frac{\delta }{{2{\Xi _j}}} > {p_j}\left( {{\btheta _{\rm{T}}}} \right)} \right\}
\end{equation}
\begin{equation}
\text{with} \quad  \eta _{j,2}^*\left( \delta  \right) \buildrel \Delta \over = \ln \frac{{1 + \frac{\delta }{{2{\Xi _j}}} - {p_j}\left( {{\btheta _{\rm{T}}}} \right)}}{{1 - {p_j}\left( {{\btheta _{\rm{T}}}} \right)}} - \left( {{p_j}\left( {{\btheta _{\rm{T}}}} \right) - \frac{\delta }{{2{\Xi _j}}}} \right)\ln \frac{{{p_j}\left( {{\btheta _{\rm{T}}}} \right)\left( {1 + \frac{\delta }{{2{\Xi _j}}} - {p_j}\left( {{\btheta _{\rm{T}}}} \right)} \right)}}{{\left( {{p_j}\left( {{\btheta _{\rm{T}}}} \right) - \frac{\delta }{{2{\Xi _j}}}} \right)\left( {1 - {p_j}\left( {{\btheta _{\rm{T}}}} \right)} \right)}}.
\end{equation}
As a result, from (\ref{P_0_Dkj_D_j}), (\ref{P_0_Dkj_D_j_1}) and (\ref{P_0_Dkj_D_j_2}), we can obtain
\begin{equation} \label{P0_Djk_Dj_F}
{\bbP_0}\left( {\left\{ {\left| {D_j^{(K)} - {D_j}} \right| \ge \frac{1}{2}\delta } \right\} \cap {{\cal F}_{j,K}}} \right) \le {e^{ - {  \eta _{j,1}\left( \delta  \right)  }K}} + {e^{ - {  \eta _{j,2}\left( \delta  \right)  }K}},
\end{equation}
where $\eta _{j,1}\left( \delta  \right)$ and $\eta _{j,2}\left( \delta  \right)$ are defined in (\ref{eta_j_1_0}) and (\ref{eta_j_2_0}), respectively.

Now, we consider the second term in (\ref{P_0_D_j_K_D_j}). From (\ref{p_j_range}) and (\ref{epsilon_range}), we know
\begin{equation}
0<{\varepsilon _j^{({\rm{L}})}} < {p_j}\left( {{\btheta _{\rm{T}}}} \right) < {\varepsilon _j^{({\rm{U}})}}<1,
\end{equation} 
and hence, by employing similar arguments, we can obtain
\begin{align} \notag
{\bbP_0}\left( {{\cal F}_{j,K}^{\rm{C}}} \right) & = {\bbP_0}\left( {\xi _j^{(K)} \notin \left[ {{\varepsilon _j^{({\rm{L}})}},{\varepsilon _j^{({\rm{U}})}}} \right]} \right)\\  \notag
&= {\bbP_0}\left( {\xi _j^{(K)} > {\varepsilon _j^{({\rm{U}})}}} \right) + {\bbP_0}\left( {\xi _j^{(K)} < {\varepsilon _j^{({\rm{L}})}}} \right)\\  \notag
&\le {\bbP_0}\left( {\sum\limits_{k = 1}^K {1 - {{\tilde u}_{jk}} - {p_j}\left( {{\btheta _{\rm{T}}}} \right)}  \ge \left( {{\varepsilon _j^{({\rm{U}})}} - {p_j}\left( {{\btheta _{\rm{T}}}} \right)} \right)K} \right) \\ \notag
& \quad + {\bbP_0}\left( {\sum\limits_{k = 1}^K {{{\tilde u}_{jk}} + {p_j}\left( {{\btheta _{\rm{T}}}} \right) - 1}  \ge \left( {{p_j}\left( {{\btheta _{\rm{T}}}} \right) - {\varepsilon _j^{({\rm{L}})}}} \right)K} \right) \\ \label{P0_F_C}
& \le {e^{ - {  \eta _{\varepsilon _j^{({\rm{U}})}}  }K}} + {e^{ - {  \eta _{\varepsilon _j^{({\rm{L}})}}  }K}},
\end{align}
where the rate functions can be expressed as
\begin{equation} \label{eta_F_1_0}
\eta _{\varepsilon _j^{({\rm{U}})}} = {\varepsilon _j^{({\rm{U}})}}\ln \frac{{{\varepsilon _j^{({\rm{U}})}}\left( {1 - {p_j}\left( {{\btheta _{\rm{T}}}} \right)} \right)}}{{{p_j}\left( {{\btheta _{\rm{T}}}} \right)\left( {1 - {\varepsilon _j^{({\rm{U}})}}} \right)}} - \ln \frac{{1 - {p_j}\left( {{\btheta _{\rm{T}}}} \right)}}{{1 - {\varepsilon _j^{({\rm{U}})}}}},
\end{equation}
\begin{equation} \label{eta_F_2_0}
\text{and} \quad \eta _{\varepsilon _j^{({\rm{L}})}}  = \ln \frac{{1 - {\varepsilon _j^{({\rm{L}})}}}}{{1 - {p_j}\left( {{\btheta _{\rm{T}}}} \right)}} - {\varepsilon _j^{({\rm{L}})}}\ln \frac{{{p_j}\left( {{\btheta _{\rm{T}}}} \right)\left( {1 - {\varepsilon _j^{({\rm{L}})}}} \right)}}{{{\varepsilon _j^{({\rm{L}})}}\left( {1 - {p_j}\left( {{\btheta _{\rm{T}}}} \right)} \right)}}.
\end{equation}
It is worth noticing that $\eta _{\varepsilon _j^{({\rm{L}})}} $ and $\eta _{\varepsilon _j^{({\rm{U}})}} $ do not depend on $\delta$. 

As a result,  from (\ref{P_0_D_j_K_D_j}), (\ref{P0_Djk_Dj_F}) and (\ref{P0_F_C}), we know that for any $j \in \{1,2,...,N+2\}$, ${\bbP_0}( {| {\hD_j^{(K)} - {D_j}} | \ge \frac{1}{2}\delta } )$ can be bounded from above as per
\begin{align} \label{P_0_D_jK_D_j_diff}
{\bbP_0}\left( {\left| {\hD_j^{(K)} - {D_j}} \right| \ge \frac{1}{2}\delta } \right) \le {e^{ - {  \eta _{j,1}\left( \delta  \right)  }K}} + {e^{ - {  \eta _{j,2}\left( \delta  \right)  }K}} + {e^{ - {  \eta _{\varepsilon _j^{({\rm{L}})}}  }K}} + {e^{ - {  \eta _{\varepsilon _j^{({\rm{U}})}}   }K}},
\end{align}
which yields an upper bound on the false alarm probability of the detector in (\ref{decision_rule}) 
\begin{align} \notag
{\bbP_0}\left( {{\varpi_j \left( \delta \right) =1}} \right) & \le \sum\limits_{i = j,N + 1,N + 2} {{e^{ - \eta _{i,1}\left( \delta  \right)K}} + {e^{ - \eta _{i,2}\left( \delta  \right)K}} + {e^{ - \eta _{\varepsilon _i^{({\rm{U}})}} K}} + {e^{ - \eta _{\varepsilon _i^{({\rm{L}})}} K}}} \\ \label{Pfa_upper_bound_proof_final_expression}
& \le 12{e^{ - \eta _j^{({0})}\left( \delta  \right)K}},
\end{align}
\begin{equation}  \label{Define_eta_j_0}
\text{with} \quad \eta _j^{({0})}\left( \delta  \right) \buildrel \Delta \over = \mathop {\min }\limits_{i = j,N + 1,N + 2} \left\{ {{\eta _{i,1}}\left( \delta  \right),{\eta _{i,1}}\left( \delta  \right),{\eta _{\varepsilon _i^{({\rm{L}})}}}, {\eta _{\varepsilon _i^{({\rm{U}})}}} } \right\}.
\end{equation}

Next, we consider the upper bound on the miss detection probability as given in (\ref{Pmd_upper_bound}).



By employing (\ref{tp}) and following the steps for obtaining (\ref{eta_j_1_0}), (\ref{eta_j_2_0}), (\ref{eta_F_1_0}), (\ref{eta_F_2_0})  and (\ref{P_0_D_jK_D_j_diff}), we can obtain 
\begin{equation} \label{P1_D_jK_D_j_tilda}
{\bbP_1}\left( {\left| {D_j^{(K)} - {{\widetilde D}_j}} \right| \ge \frac{1}{2}\delta } \right) \le {e^{ - {{\tilde \eta }_{j,1}}\left( \delta  \right)K}} + {e^{ - {{\tilde \eta }_{j,2}}\left( \delta  \right)K}} + {e^{ - {{\tilde \eta }_{\varepsilon _j^{({\rm{L}})}}} K}} + {e^{ - {{\tilde \eta }_{\varepsilon _j^{({\rm{U}})}}} K}}
\end{equation}
where ${{\tilde \eta }_{j,1}}(\delta)$, ${{\tilde \eta }_{j,2}}(\delta)$, ${{\tilde \eta }_{\varepsilon _j^{({\rm{L}})}}}$ and ${{\tilde \eta }_{\varepsilon _j^{({\rm{U}})}}}$ can be expressed as
\begin{equation} \label{eta_j_1_0_tilda}
\tilde{\eta} _{j,1}\left( \delta  \right) =  \tilde{\eta} _{j,1}^*\left( \delta  \right)\bone\left\{ {\frac{\delta }{{2{\Xi _j}}} \le 1 - {\tp_j}\left( {{\btheta _{\rm{T}}}} \right)} \right\} + \infty \bone\left\{ {\frac{\delta }{{2{\Xi _j}}} > 1 - {\tp_j}\left( {{\btheta _{\rm{T}}}} \right)} \right\},
\end{equation}
\begin{equation} \label{eta_j_2_0_tilda}
\tilde{\eta} _{j,2}\left( \delta  \right) =  \tilde{\eta} _{j,2}^*\left( \delta  \right){\bone}\left\{ {\frac{\delta }{{2{\Xi _j}}} \le {\tp_j}\left( {{\btheta _{\rm{T}}}} \right)} \right\} + \infty {\bone}\left\{ {\frac{\delta }{{2{\Xi _j}}} > {\tp_j}\left( {{\btheta _{\rm{T}}}} \right)} \right\},
\end{equation}
\begin{equation} \label{eta_F_2_0_tilda}
\tilde{\eta} _{\varepsilon _j^{({\rm{L}})}} = \ln \frac{{1 - {\varepsilon _j^{({\rm{L}})}}}}{{1 - {\tp_j}\left( {{\btheta _{\rm{T}}}} \right)}} - {\varepsilon _j^{({\rm{L}})}}\ln \frac{{{\tp_j}\left( {{\btheta _{\rm{T}}}} \right)\left( {1 - {\varepsilon _j^{({\rm{L}})}}} \right)}}{{{\varepsilon _j^{({\rm{L}})}}\left( {1 - {\tp_j}\left( {{\btheta _{\rm{T}}}} \right)} \right)}}.
\end{equation}
\begin{equation} \label{eta_F_1_0_tilda}
\tilde{\eta}  _{\varepsilon _j^{({\rm{U}})}}  = {\varepsilon _j^{({\rm{U}})}}\ln \frac{{{\varepsilon _j^{({\rm{U}})}}\left( {1 - {\tp_j}\left( {{\btheta _{\rm{T}}}} \right)} \right)}}{{{\tp_j}\left( {{\btheta _{\rm{T}}}} \right)\left( {1 - {\varepsilon _j^{({\rm{U}})}}} \right)}} - \ln \frac{{1 - {\tp_j}\left( {{\btheta _{\rm{T}}}} \right)}}{{1 - {\varepsilon _j^{({\rm{U}})}}}},
\end{equation}
and $\tilde{\eta} _{j,1}^*\left( \delta  \right)$ and $\tilde{\eta} _{j,2}^*\left( \delta  \right)$ are defined as
\begin{equation}  \label{eta_j_result_tilda}
\tilde{\eta} _{j,1}^*\left( \delta  \right) \triangleq \left( {\frac{\delta }{{2{\Xi _j}}} + {\tp_j}\left( {{\btheta _{\rm{T}}}} \right)} \right)\ln \frac{{\left( {\frac{\delta }{{2{\Xi _j}}} + {\tp_j}\left( {{\btheta _{\rm{T}}}} \right)} \right)\left( {1 - {\tp_j}\left( {{\btheta _{\rm{T}}}} \right)} \right)}}{{{\tp_j}\left( {{\btheta _{\rm{T}}}} \right)\left( {1 - \frac{\delta }{{2{\Xi _j}}} - {\tp_j}\left( {{\btheta _{\rm{T}}}} \right)} \right)}} - \ln \frac{{1 - {\tp_j}\left( {{\btheta _{\rm{T}}}} \right)}}{{1 - \frac{\delta }{{2{\Xi _j}}} - {\tp_j}\left( {{\btheta _{\rm{T}}}} \right)}},
\end{equation}
\begin{equation}
\tilde{\eta}_{j,2}^*\left( \delta  \right) \buildrel \Delta \over = \ln \frac{{1 + \frac{\delta }{{2{\Xi _j}}} - {\tp_j}\left( {{\btheta _{\rm{T}}}} \right)}}{{1 - {\tp_j}\left( {{\btheta _{\rm{T}}}} \right)}} - \left( {{\tp_j}\left( {{\btheta _{\rm{T}}}} \right) - \frac{\delta }{{2{\Xi _j}}}} \right)\ln \frac{{{\tp_j}\left( {{\btheta _{\rm{T}}}} \right)\left( {1 + \frac{\delta }{{2{\Xi _j}}} - {\tp_j}\left( {{\btheta _{\rm{T}}}} \right)} \right)}}{{\left( {{\tp_j}\left( {{\btheta _{\rm{T}}}} \right) - \frac{\delta }{{2{\Xi _j}}}} \right)\left( {1 - {\tp_j}\left( {{\btheta _{\rm{T}}}} \right)} \right)}}.
\end{equation}
Moreover, noticing that
\begin{equation}
{{\bbP_1}\left( {\left| {\hD_{N + i}^{(K)} - {D_{N + i}}} \right| \ge \frac{1}{2}\delta } \right)} = {{\bbP_0}\left( {\left| {\hD_{N + i}^{(K)} - {D_{N + i}}} \right| \ge \frac{1}{2}\delta } \right)}, \;   i=1,2,
\end{equation}
by employing (\ref{Pmd_upper_bound}), (\ref{P_0_D_jK_D_j_diff}) and (\ref{P1_D_jK_D_j_tilda}), we can obtain
\begin{align} \notag
{\bbP_1}\left( {{\varpi_j \left( \delta \right) =0}} \right) & \le {e^{ - {{\tilde \eta }_{j,1}}\left( \delta  \right)K}} + {e^{ - {{\tilde \eta }_{j,2}}\left( \delta  \right)K}} + {e^{ - {{\tilde \eta }_{\varepsilon _j^{({\rm{L}})}}} K}} + {e^{ - {{\tilde \eta }_{\varepsilon _j^{({\rm{U}})}}} K}} \\  \notag
& \quad + \sum\limits_{i = N + 1}^{N + 2} {{e^{ - \eta _{i,1}\left( \delta  \right)K}} + {e^{ - \eta _{i,2}\left( \delta  \right)K}} + {e^{ - \eta _{\varepsilon _i^{({\rm{U}})}} K}} + {e^{ - \eta _{\varepsilon _i^{({\rm{L}})}} K}}} \\ \label{Pmd_upper_bound_proof_final_expression}
& \le 12{e^{ - \eta _j^{({1})}\left( \delta  \right)K}},
\end{align}
where the quantity $\eta _j^{({1})}\left( \delta  \right) $ is defined as
\begin{equation} \label{Define_eta_j_1}
\eta _j^{({1})}\left( \delta  \right) \buildrel \Delta \over = \mathop {\min }\limits_{i = N + 1,N + 2} \left\{ {{{\tilde \eta }_{j,1}}\left( \delta  \right),{{\tilde \eta }_{j,2}}\left( \delta  \right),{\eta _{i,1}}\left( \delta  \right),{{\tilde \eta }_{\varepsilon _j^{({\rm{L}})}}},{{\tilde \eta }_{\varepsilon _j^{({\rm{U}})}}},{\eta _{i,1}}\left( \delta  \right),{\eta _{\varepsilon _i^{({\rm{L}})}}},{\eta _{\varepsilon _i^{({\rm{U}})}}}} \right\}.
\end{equation}
\end{IEEEproof}

As demonstrated by {Theorem \ref{Theorem_Performance}}, the false alarm and miss probabilities of the proposed detector in (\ref{decision_rule}) are guaranteed to decay exponentially as $K$ increases. The decay rates are illustrated in (\ref{Define_eta_j_0}) and (\ref{Define_eta_j_1}) which depend on the choice of $\delta$. In general, a smaller $\delta$ leads to a larger false alarm probability and a smaller miss probability. Hence, the trade-off between the false alarm and miss probabilities can be sought via altering the value of $\delta$. 

Using {Theorem \ref{Theorem_Performance}}, the  average detector error probability  $P_{\rm{e}}$ can be bounded from above as per
\begin{align} \notag
{P_{{\rm{e}}}} & = \frac{1}{N}\sum\limits_{j \in {\cal U}} {{\bbP_0}\left( {{\varpi _j} = 1} \right)}  + \frac{1}{N}\sum\limits_{j \in {\cal V}} {{\bbP_1}\left( {{\varpi _j} = 0} \right)} \\ \notag
& \le \frac{{12}}{N}\sum\limits_{j \in {\cal U}} {{e^{ - \eta _j^{({\rm{0}})}\left( \delta  \right)K}}}  + \frac{{12}}{N}\sum\limits_{j \in {\cal V}} {{e^{ - \eta _j^{(1)}\left( \delta  \right)K}}} \\
& \le {C_{{\rm{e}}}}{e^{ - {\eta _{{\rm{e}}}}\left( \delta  \right)K}},
\end{align}
where the positive constants $C_{\rm{e}}$ and $\eta_{\rm{e}}(\delta)$ are defined as
\begin{equation}
{C_{{\rm{e}}}} = 12 \quad \text{and} \quad  {\eta _{{\rm{e}}}}\left( \delta  \right) \buildrel \Delta \over = \mathop {\min }\limits_{j = 1,2,...,N} \left\{ {\eta _j^{({\rm{0}})}\left( \delta  \right),\eta _j^{(1)}\left( \delta  \right)} \right\}.
\end{equation} 
This observation is summarized in the following corollary.
%
%
\begin{corollary} \label{Corollary_Percentage_Misclassified}
	If (\ref{lemma_delta_constraint}) holds,
	then the average detector error probability decreases at least  exponentially  as $K$ increases.
\end{corollary}

It is worth pointing out that the sufficient condition on $\delta$ in  {Theorem \ref{Theorem_Performance}} and  {Corollary \ref{Corollary_Percentage_Misclassified}}
 are generally not necessary, which is observed in all the numerical experiments that we carried out. 

\section{Simulation Results}
\label{Section_Numerical}

In this section, we first introduce how to implement the proposed attack detector in practice, and then we test the performance of the proposed attack detector to corroborate the theoretical results in previous sections. 

\subsection{Implementation of the Attack Detector}


By employing (\ref{Definition_hD_jK}), $\hD_{j}^{(K)}$, $\hD_{N+1}^{(K)}$ and $\hD_{N+2}^{(K)}$ can be computed, and thereby the analytical expression of ${\cC}( \btheta_j, \hD_{j}^{(K)}  )$ can be obtained. Note that every point $\btheta$ in the common area of ${\cal R}=( {{\btheta _{N+1}},\hD_{N+1}^{(K)}}, \delta )$ and ${\cal R}( {{\btheta _{N+2}},\hD_{N+2}^{(K)}}, \delta )$ satisfies the condition
\begin{equation} \label{Condition_implementation}
\left\{ \begin{array}{l}
 - \delta  \le \left\| {\btheta  - {\btheta _{N + 1}}} \right\| - {\widehat D}_{N + 1}^{\left( K \right)} \le \delta, \\
 - \delta  \le \left\| {\btheta  - {\btheta _{N + 2}}} \right\| - {\widehat D}_{N + 2}^{\left( K \right)} \le \delta.
\end{array} \right.
\end{equation}
Therefore, to implement the attack detector in (\ref{decision_rule}), we only need to check whether any point on the circle ${\cC}( \btheta_j, \hD_{j}^{(K)}  )$ satisfies the condition in (\ref{Condition_implementation}) or not. 

One brute force way to do this is to discretize ${\cC}( \btheta_j, \hD_{j}^{(K)}  )$  to finitely many points which are evenly spaced along the circle, and then we check the condition in (\ref{Condition_implementation}) for these points. In particular, we can discretize ${\cC}( \btheta_j, \hD_{j}^{(K)}  )$ to $M$ points $\{ \btheta_{\cC}^{(m)} \}_{m=1}^M$ in the way that
\begin{equation} \label{Discretize_circle}
\btheta_{\cC}^{(m)} = \lsb x_j+ \hD_{j}^{(K)} \cos\left( \frac{2\pi}{M}(m-1) \right),  y_j+ \hD_{j}^{(K)} \sin\left( \frac{2\pi}{M}(m-1) \right) \rsb,
\end{equation}
where $x_j$ and $y_j$ are the coordinates of the $j$-th sensor.
We summarize this implementation in Algorithm \ref{Algorithm_Implementation}. Intuitively, we expect that this approach may not work well for small $M$.

\begin{algorithm}[htb]
	\caption{Implementation of attack detector}
	\begin{algorithmic}[1]
		\STATE \textbf{Input}: $\{u_{ik}\}_{k=1}^K$ for $i=N+1, N+2$, $\{\tu_{jk}\}_{k=1}^K$, and $\delta$;
		\STATE  \textbf{Output}: $\varpi_j(\delta)$;
		\STATE Compute  $\hD_{j}^{(K)}$, $\hD_{N+1}^{(K)}$ and $\hD_{N+2}^{(K)}$ by employing (\ref{Definition_hD_jK});
		\STATE Discretize ${\cC}( \btheta_j, \hD_{j}^{(K)}  )$ to $\{ \btheta_{\cC}^{(m)} \}_{m=1}^M$ by employing (\ref{Discretize_circle});
		\STATE  $m \leftarrow 1$ \textbf{and} $\varpi_j(\delta) \leftarrow 1$;
		\WHILE{$m \le M$ \textbf{and} $\varpi_j(\delta)=1$ }
		\IF{$\btheta_{\cC}^{(m)} \in \cS$ \textbf{and}  (\ref{Condition_implementation}) holds} 
		\STATE $\varpi_j(\delta) \leftarrow 0$;
		\ENDIF
		\STATE $m \leftarrow m+1$;
		\ENDWHILE
	\end{algorithmic}
	\label{Algorithm_Implementation}
\end{algorithm}

\subsection{Simulation Setup and Results}

\begin{figure}
	\centerline{
		\includegraphics[height=0.5\textwidth]{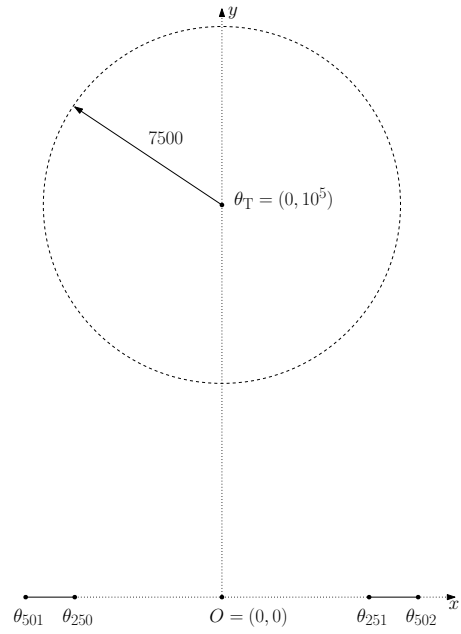}
	}
	\caption{Simulation setup.
	}	
	\label{Fig_Simulation_Setup}	
\end{figure}


\begin{figure}
	\centerline{
		\includegraphics[width=0.68\textwidth]{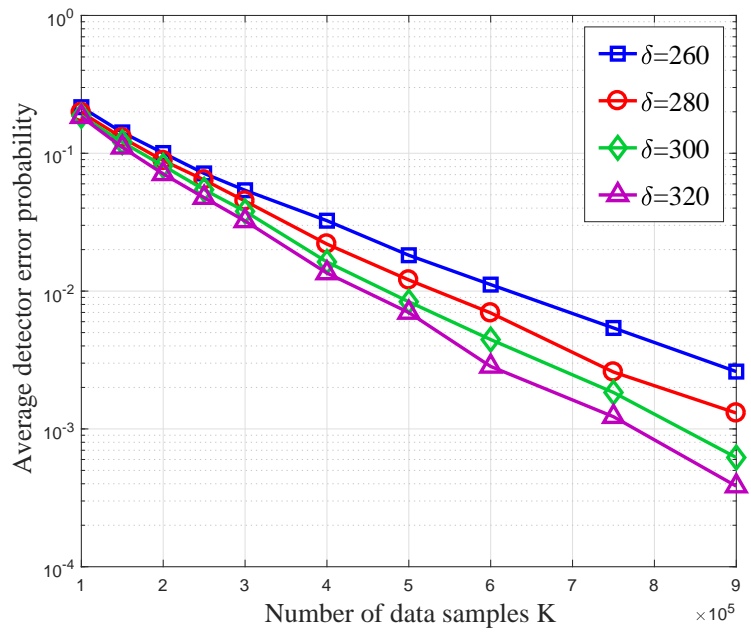}
	}
	\caption{Attack identification performance of the proposed detectors.
	}	
	\label{Fig_Percentage_105}	
\end{figure}

\begin{figure}
	\centerline{
		\includegraphics[width=0.68\textwidth]{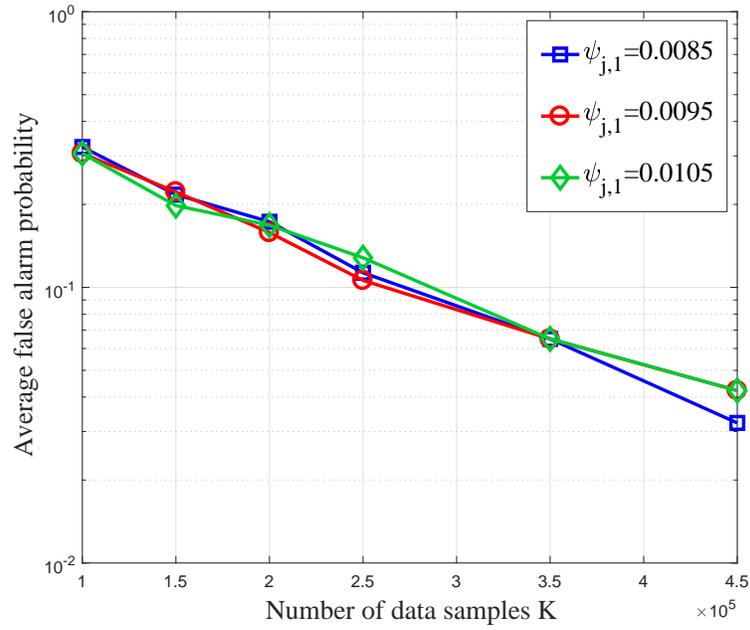}
	}
	\caption{Average false alarm probabilities of the proposed detector under different attacks.
	}	
	\label{Fig_false_alarm}	
\end{figure}

\begin{figure}
	\centerline{
		\includegraphics[width=0.68\textwidth]{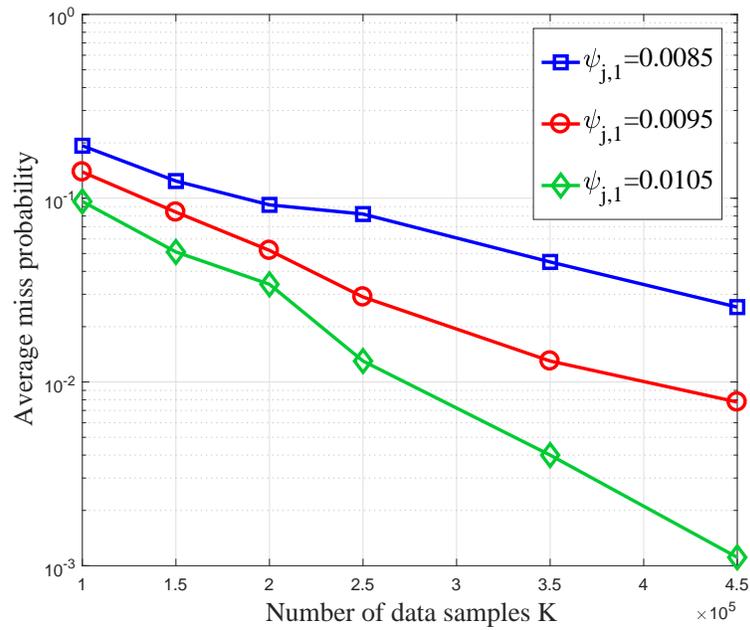}
	}
	\caption{Average miss probabilities of the proposed detector under different attacks.
	}	
	\label{Fig_miss}	
\end{figure}

\begin{figure}
	\centerline{
		\includegraphics[width=0.68\textwidth]{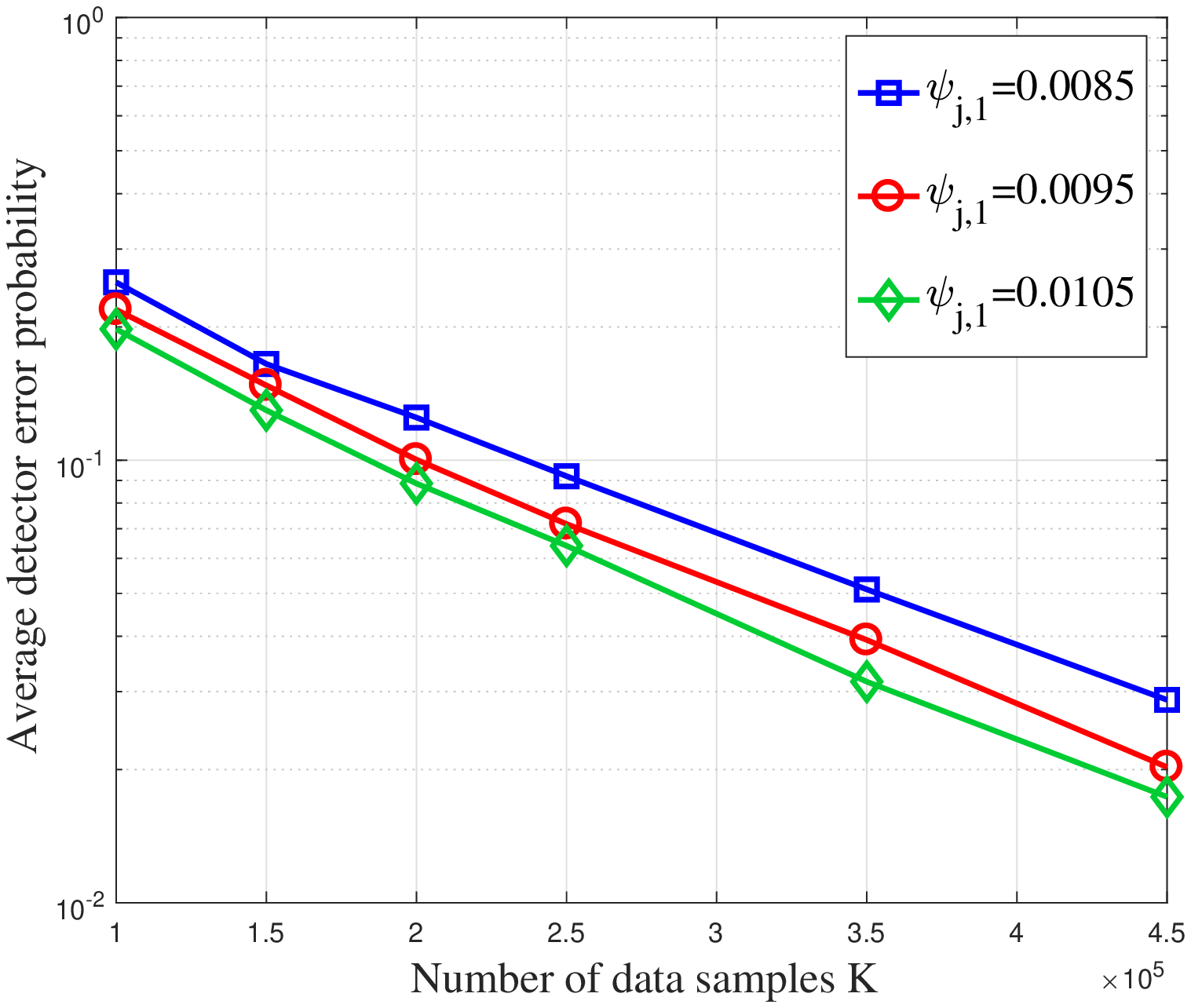}
	}
	\caption{Attack identification performance of the proposed detector under different attacks.
	}	
	\label{Fig_Percentage_different_attacks}	
\end{figure}

The simulation setup is illustrated in Fig. \ref{Fig_Simulation_Setup}. Consider a sensor network consisting of two groups of sensors with $N=500$. The two secure sensors are located at ${\btheta}_{501} = (-10^3,0)$ and ${\btheta}_{502} = (10^3,0)$, respectively. The rest of sensors are all located along the $x$-axis, and are partitioned into two groups. The first group of sensors 
$\{1, 2,...,250, 501\}$  are evenly spaced between $ (-10^3,0)$ and $ (-0.9\times10^3,0)$, while the second group of sensors 
$\{251 , 252, ..., 500, 502 \}$are evenly spaced between  $(0.9\times10^3,0)$ and $ (10^3,0)$.  The ROI $\cA$ is a disc centered at $(0, 10^5)$ and with radius equal to $7500$. The target is located at ${\btheta}_{\rm{T}}=(0, 10^5)$. In the simulation, $P_0=1$, $D_0=10^5$, and $\gamma=2$. When employing Algorithm \ref{Algorithm_Implementation} to implement the attack detector, $M$ is chosen to be $2\times 10^5$. In addition, the threshold $\tau_j =1$ for all $j$, and $n_{jk}$ follows a Gaussian distribution with zero mean and unit variance. We assume that $250$ sensors
 $\{ 1,2,...,250 \}$ are under the MiMA as described in (\ref{man_in_the_middle_attack_Rule}) with $\psi_{j,0}=0$ and $\psi_{j,1}=0.0105$ for $j=1,2,...,250$. 
The average detector error probability over $900$ Monte Carlo runs versus the number $K$ of data samples are depicted on a
log scale in Fig. \ref{Fig_Percentage_105} for four detectors with $\delta=260,280,300,320$, respectively. It is seen from Fig. \ref{Fig_Percentage_105} that for each detector, the average detector error probability decreases exponentially as $K$ grows which agrees with the theoretical results in the previous section. Moreover, as illustrated in Fig. \ref{Fig_Percentage_105}, the larger the value of $\delta$, the better the identification performance, which implies that in this simulation, as $\delta$ increases, the gain obtained in the false alarm probability is larger than the loss in the miss probability, since the number of attacked sensors is the same as the number of unattacked sensors.


Now, we consider the attack identification performance of the proposed detector under different attacks. 
In Fig. \ref{Fig_false_alarm}, Fig. \ref{Fig_miss} and Fig. \ref{Fig_Percentage_different_attacks}, the simulation setup is the same as that for Fig. \ref{Fig_Percentage_105}, except $\delta=280$.
The different attacks are all MiMA and for $j=1,2,...,250$, $[\psi_{j,0}, \psi_{j,1}]=[0, 0.0085]$, $[0, 0.0095]$, and $[0, 0.0105]$, respectively.
It is seen from Fig. \ref{Fig_false_alarm} that under different attacks, the false alarm probabilities achieved by the detector are very close, which agrees with the fact that the false alarm probability does not depend on the attacks, but is only determined by $\delta$.
As expected from the intuition that the attack which brings about a larger impact on the statistical model of the data should be easier to be detected, Fig. \ref{Fig_miss} demonstrates that the larger the value of $\psi_{j,1}$, the smaller the average miss probability.
Fig. \ref{Fig_Percentage_different_attacks} also corroborates the intuition that the larger the attack impact on the statistical model of the data, the better the attack identification performance that can be achieved.


\section{Conclusions}
\label{Section_Conclusion}

This work has investigated the attack detection in sensor network target localization systems with quantized data. By exploring the impact of the attacks on  the statistical model of the sensor data, we have revealed that from the perspective of the NMLE,  the essential effect of attacks is a falsification of the estimated distance between the target and each attacked sensor,
and hence, gives rise to a geometric inconsistency among the attacked  and  unattacked sensors. Motivated by this fact,  a class of detectors are proposed to detect the attacks in the sensor network via scrutinizing the existence of the geometric inconsistency. A rigorous detection performance analysis for the proposed detectors has been carried out, showing that the false alarm and miss probabilities decay exponentially  as the number of data samples at each sensor grows, which implies that 
for a sufficiently large number of samples, the proposed detectors can identify the attacked sensors with any required level of accuracy.

\appendices

\section{Proof of Lemma \ref{Lemma_overlap_area}} \label{Proof_Lemma_overlap_area}

\begin{figure}[htb]
	\centerline{
		\includegraphics[width=0.5\textwidth]{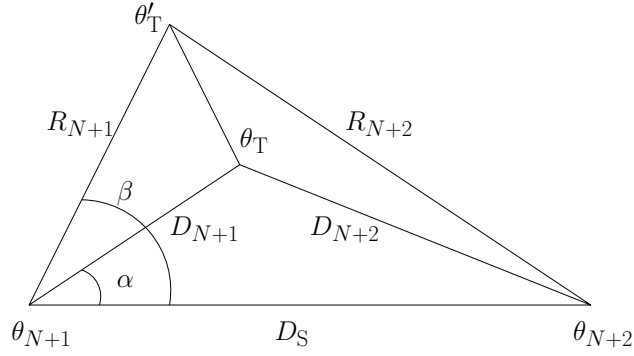}
	}
	\caption{Geometric illustration.
	}	
	\label{Fig_Geometry_fig}	
\end{figure}

Consider $R_{N+1}$ and $R_{N+2}$ which satisfy
\begin{equation} \label{R_i_D_i}
|R_{i} - D_{i}| \le \delta  < \Upsilon, \; \text{ for } i=N+1, N+2,
\end{equation} 
and denote
\begin{equation}
{\btheta_\teT '} \triangleq \cC(\btheta_{N+1}, R_{N+1}) \cap \cC(\btheta_{N+2}, R_{N+2}).
\end{equation}
From (\ref{assumption_D_N_1_N_2}) and (\ref{R_i_D_i}), we know that
\begin{equation}
R_{N+1} + R_{N+2} \ge D_{N+1} + D_{N+2} - 2\delta > \mathop {\inf }\limits_{{\btheta_\teT} \in {\cA}} \{ D_{N+1} + D_{N+2}\} - 2\Upsilon_2 = D_\teS,
\end{equation}
and moreover, by employing (\ref{assumption_Area_Widely_Separated}) and (\ref{R_i_D_i}), we can obtain
\begin{equation}
|R_{N+1} - R_{N+2}| < |D_{N+1} - D_{N+2}| + 2\delta < D_\teU - D_\teL +2\Upsilon \le D_\teU - D_\teL +2\Upsilon_1 < D_\teS.
\end{equation}
Thus, $R_{N+1}$, $R_{N+2}$ and $D_\teS$ can be the sides of a triangle, and hence,  ${\btheta_\teT '}$ exists and cannot be on the line passing through $\btheta_{N+1}$ and $\btheta_{N+2}$, which implies that the angle $\beta \triangleq \angle {\btheta_{\teT}'}{\btheta_{N+1}}{\btheta_{N+2}}$ in Fig. \ref{Fig_Geometry_fig} satisfies $\beta \in (0, \pi)$.

Let $\alpha$ denote the angle $\angle {\btheta_{\teT}}{\btheta_{N+1}}{\btheta_{N+2}}$ as illustrated in Fig. \ref{Fig_Geometry_fig}. By the law of cosines, we can obtain
\begin{equation} \label{Definition_d}
d(R_{N+1}, R_{N+2}) \triangleq \|  \btheta_\teT ' - \btheta_\teT \| = R_{N+1}^2 + D_{N+1}^2 - 2R_{N+1}D_{N+1}\cos(\beta -\alpha).
\end{equation}
According to  {Assumption \ref{Assumption_Area}}, we know
\begin{equation}
{D_{\rm{S}}} > \left| {{D_{N + 1}} - {D_{N + 2}}} \right| \quad \text{and} \quad {D_{N + 1}} + {D_{N + 2}} > {D_{\rm{S}}},
\end{equation}
which yields $\alpha \in (0,\pi)$, 
and hence,
\begin{equation}
\beta - \alpha \in (-\pi, \pi).
\end{equation}
From (\ref{Definition_d}), we know that for any given $R_{N+1}$, $d(R_{N+1}, R_{N+2})$ is maximized when $\cos(\beta -\alpha)$ is minimized. Since $\alpha$ is fixed and $\beta - \alpha \in (-\pi, \pi)$, $\cos(\beta -\alpha)$ is minimized when $\beta$ is either maximized or minimized, which implies that $d(R_{N+1}, R_{N+2})$ is maximized when $\beta$ is either maximized or minimized.

Furthermore, by the law of cosines, we can obtain
\begin{equation}
\cos(\beta) = \frac{R_{N+1}^2+D_\teS^2-R_{N+2}^2}{2R_{N+1}D_\teS}.
\end{equation}
Since $\beta \in (0,\pi)$ and $\cos(\beta) $ is decreasing over $\beta \in (0,\pi)$, for any given $R_{N+1}$, $\beta$ is maximized if $R_{N+2}$ is maximized, while $\beta$ is minimized if $R_{N+2}$ is minimized. Therefore, for any given $R_{N+1}$, $d(R_{N+1}, R_{N+2})$ is maximized only when $R_{N+2} = D_{N+2} +\delta$ or $R_{N+2} = D_{N+2} -\delta$, since $|R_{N+2} - D_{N+2}| \le \delta$. 

Similarly, for any given $R_{N+2}$, $d(R_{N+1}, R_{N+2})$ is maximized only when $R_{N+1} = D_{N+1} + \delta$ or $R_{N+1} = D_{N+1} - \delta$. 

Thus, for any given $R_{N+1}$ and $R_{N+2}$ satisfying (\ref{R_i_D_i}), the maximal $d(R_{N+1}, R_{N+2})$ can only be achieved when $R_{N+1} \in \{ D_{N+1} -\delta,  D_{N+1} +\delta\}$ and $R_{N+2} \in \{ D_{N+2} -\delta,  D_{N+2} +\delta\}$. To this end, in order to prove ${\mathop  \cap_{i =1}^{2} {\cal R}\left( {{\btheta _{N+i}},\hD_{N+i}^{(K)}}, \delta \right)}  \subseteq \cB(\btheta_\teT, \Phi(\delta))$, we only need to consider 
\begin{align} \label{R_Nplus1_range}
R_{N+1} & \in \{ D_{N+1} -\delta,  D_{N+1} +\delta\}, \\ \label{R_Nplus2_range}
R_{N+2} & \in \{ D_{N+2} -\delta,  D_{N+2} +\delta\},
\end{align}
and show ${\btheta}_{\teT}' \in \cB(\btheta_\teT, \Phi(\delta))$.

Without loss of generality, we assume that $\btheta_{N+1} = {\bf 0}$, $\btheta_{N+2} = (D_\teS,0)$, and $\btheta_\teT$ is in the half space above the line passing through $\btheta_{N+1} $ and $\btheta_{N+2}$. Since ${\btheta_\teT} \triangleq \cC(\btheta_{N+1}, D_{N+1}) \cap \cC(\btheta_{N+2}, D_{N+2})$, we can obtain
\begin{equation}
\left\{ \begin{array}{l}
x_\teT^2 + y_\teT^2 = D_{N + 1}^2,\\
{\left( {{x_\teT} - {D_\teS}} \right)^2} + y_\teT^2 = D_{N + 2}^2,
\end{array} \right.
\end{equation}
which yields
\begin{equation}  \label{x_T}
\left\{ \begin{array}{l}
x_\teT = \frac{D_{N + 1}^2 - D_{N + 2}^2 + D_\teS^2}{2D_\teS}, \\
y_\teT = \sqrt{D_{N + 1}^2 - \lp \frac{D_{N + 1}^2 - D_{N + 2}^2 + D_\teS^2}{2D_\teS}  \rp^2}.
\end{array} \right.
\end{equation}
Similarly, with regard to ${\btheta_\teT '} = (x_\teT', y_\teT') = \cC(\btheta_{N+1}, R_{N+1}) \cap \cC(\btheta_{N+2}, R_{N+2})$, we also can obtain
\begin{equation} \label{x_T_prime}
\left\{ \begin{array}{l}
x_\teT' = \frac{R_{N + 1}^2 - R_{N + 2}^2 + D_\teS^2}{2D_\teS}, \\
y_\teT' = \sqrt{R_{N + 1}^2 - \lp \frac{R_{N + 1}^2 - R_{N + 2}^2 + D_\teS^2}{2D_\teS}  \rp^2}.
\end{array} \right.
\end{equation}

By employing (\ref{x_T}) and (\ref{x_T_prime}),  $d{( {{R_{N + 1}}, {R_{N + 2}}} )^2}$ can be expressed as
\begin{align} \notag
& d{\left( {{R_{N + 1}}, {R_{N + 2}}} \right)^2} \\ \notag
& = \underbrace {{{\left( {\frac{{R_{N + 1}^2 - R_{N + 2}^2 + D_\teS^2}}{{2{D_\teS}}} - \frac{{D_{N + 1}^2 - D_{N + 2}^2 + D_\teS^2}}{{2{D_\teS}}}} \right)}^2}}_{{d_x}} \\ \label{d_x_d_y}
& \quad + \underbrace {{{\left[ {\sqrt {R_{N + 1}^2 - {{\left( {\frac{{R_{N + 1}^2 - R_{N + 2}^2 + D_\teS^2}}{{2{D_\teS}}}} \right)}^2}}  - \sqrt {D_{N + 1}^2 - {{\left( {\frac{{D_{N + 1}^2 - D_{N + 2}^2 + D_\teS^2}}{{2{D_\teS}}}} \right)}^2}} } \right]}^2}}_{{d_y}}.
\end{align}
From (\ref{R_Nplus1_range}) and (\ref{R_Nplus2_range}), $d_x$ can be bounded from above as per
\begin{align} \notag
{d_x} & = {\left( {\frac{{R_{N + 1}^2 - D_{N + 1}^2 + D_{N + 2}^2 - R_{N + 2}^2}}{{2{D_\teS}}}} \right)^2}\\ \notag
& = {\left( {\frac{{\left( {{R_{N + 1}} - {D_{N + 1}}} \right)\left( {{R_{N + 1}} + {D_{N + 1}}} \right) + \left( {{D_{N + 2}} - {R_{N + 2}}} \right)\left( {{D_{N + 2}} + {R_{N + 2}}} \right)}}{{2{D_\teS}}}} \right)^2}\\ \notag
& \le \frac{1}{{D_\teS^2}}{\delta^2}{\left( {{D_{N + 1}} + {D_{N + 2}} + \delta} \right)^2}\\ \label{d_x_upperbound}
& \le \frac{1}{{D_\teS^2}}{\left( {2{D_\teU} + \delta} \right)^2}{\delta^2}
\end{align}
and moreover, by using the fact that $\sqrt{|x|} - \sqrt{|y|} \le \sqrt{|x-y|}$ for any $x$ and $y$, $d_y$ can be bounded from above as per
\begin{align} \notag
{d_y} & \le \left| {R_{N + 1}^2 - {{\left( {\frac{{R_{N + 1}^2 - R_{N + 2}^2 + D_\teS^2}}{{2{D_\teS}}}} \right)}^2} - D_{N + 1}^2 + {{\left( {\frac{{D_{N + 1}^2 - D_{N + 2}^2 + D_\teS^2}}{{2{D_\teS}}}} \right)}^2}} \right|\\  \notag
&  \le \left| {\left( {{R_{N + 1}} - {D_{N + 1}}} \right)\left( {{R_{N + 1}} + {D_{N + 1}}} \right)} \right| \\ \label{temp_d_y}
& \quad + \left| {\frac{{\left( {R_{N + 1}^2 - R_{N + 2}^2 - D_{N + 1}^2 + D_{N + 2}^2} \right)\left( {R_{N + 1}^2 - R_{N + 2}^2 + D_{N + 1}^2 - D_{N + 2}^2 + 2D_\teS^2} \right)}}{{4D_\teS^2}}} \right|.
\end{align}
By employing  (\ref{R_Nplus1_range}), (\ref{R_Nplus2_range}) and (\ref{temp_d_y}), we can obtain
\begin{align} \notag
{d_y} & \le \delta \left( {2{D_{N + 1}} + r} \right) + \frac{{\left| {R_{N + 1}^2 - D_{N + 1}^2} \right| + \left| {R_{N + 2}^2 - D_{N + 2}^2} \right|}}{{4D_\teS^2}}\\ \notag
& \quad \times \left( {\left| {R_{N + 1}^2 - D_{N + 2}^2} \right| + \left| {R_{N + 2}^2 - D_{N + 1}^2} \right| + 2D_\teS^2} \right)\\ \notag
& \le  \delta \left( {2{D_{N + 1}} +  \delta } \right) + \frac{{ \delta \left( {2{D_{N + 1}} +  \delta } \right) +  \delta \left( {2{D_{N + 2}} +  \delta } \right)}}{{4D_\teS^2}}\\ \notag
& \quad \times \left( {\left| {\left( {{R_{N + 1}} - {D_{N + 2}}} \right)\left( {{R_{N + 1}} + {D_{N + 2}}} \right)} \right| + \left| {\left( {{R_{N + 2}} - {D_{N + 1}}}  \right)\left( {{R_{N + 2}} + {D_{N + 1}}} \right)} \right| + 2D_\teS^2} \right)\\  \notag
& \le  \delta \left( {2{D_{N + 1}} +  \delta } \right) + \frac{{ \delta \left( {{D_{N + 1}} + {D_{N + 2}} +  \delta } \right) }}{{D_\teS^2}}\\  \label{d_y_temp1}
& \quad  \times \left[ {\left( {\left| {{D_{N + 1}} - {D_{N + 2}}} \right| +  \delta } \right)\left( {{D_{N + 1}} + {D_{N + 2}} +  \delta } \right) + D_\teS^2} \right] \\  \notag
& \le  \delta \left( {2{D_{N+1}} +  \delta  } \right) +  \delta \left( {2{D_\teU} +  \delta  } \right) \left[ {\frac{{\left( {{D_\teU} - {D_\teL} +  \delta } \right)\left( {{D_{N + 1}} + {D_{N + 2}} +  \delta } \right)}}{{D_\teS^2}} + 1} \right]\\ \label{d_y_temp2}
& \le  \delta \left( {2{D_\teU} +  \delta  } \right) +  \delta \left( {2{D_\teU} +  \delta } \right)\left( {\frac{{2{D_\teU} +  \delta  }}{{{D_\teS}}} + 1} \right)\\ \label{d_y_upperdound}
&  \le \left( {2{D_\teU} +  \delta  } \right)\left( {\frac{{2{D_\teU} +  \delta  }}{{{D_\teS}}} + 2} \right) \delta ,
\end{align}
where (\ref{d_y_temp1}) is from (\ref{R_i_D_i}), and (\ref{d_y_temp2}) is due to $D_{N+1} \le D_\teU$ and {Assumption \ref{Assumption_Area}} that $D_\teS > D_\teU -D_\teL + 2 \Upsilon > D_\teU -D_\teL +  \delta $, since $ \delta < \Upsilon$.

From (\ref{d_x_d_y}), (\ref{d_x_upperbound}) and (\ref{d_y_upperdound}), we can obtain
\begin{align} \notag
d{\left( {{R_{N + 1}},{R_{N + 2}}} \right)^2} & \le \frac{1}{{D_S^2}}{\left( {2{D_U} + \delta} \right)^2}{ \delta^2} + \left( {2{D_U} + \delta} \right)\left( {\frac{{2{D_U} + \delta}}{{{D_S}}} + 2} \right)\delta\\ \notag
& \le \left( {2{D_U} + \delta} \right)\left[ {\frac{{2{D_U} + \delta}}{{{D_S}}}\left( {\frac{\delta}{{{D_S}}} + 1} \right) + 2} \right]\delta\\
& < \left( {2{D_U} + \Upsilon } \right)\left[ {\frac{{2{D_U} + \Upsilon }}{{{D_S}}}\left( {\frac{\Upsilon }{{{D_S}}} + 1} \right) + 2} \right]\delta,
\end{align}
which implies
\begin{equation}
d{\left( {{R_{N + 1}},{R_{N + 2}}} \right)} <  \left( {2{D_U} + \Upsilon } \right)^{\frac{1}{2}}\left[ {\frac{{2{D_U} + \Upsilon }}{{{D_S}}}\left( {\frac{\Upsilon }{{{D_S}}} + 1} \right) + 2} \right]^{\frac{1}{2}}\sqrt{\delta},
\end{equation}
and therefore,
\begin{equation}
{\btheta}_{\teT}' \in \cB(\btheta_\teT, \Phi(\delta)).
\end{equation}

Moreover, note that $\cB(\btheta_\teT, \delta) \subset \cR(\btheta_{N+1}, D_{N+1}, \delta)$ and  $\cB(\btheta_\teT, \delta) \subset \cR(\btheta_{N+2}, D_{N+2}, \delta)$, and hence $\cB(\btheta_\teT, \delta) \subseteq {{  \cap_{i =1}^{2} {\cal R}\left( {{\btheta _{N+i}},\hD_{N+i}^{(K)}}, \delta \right)} }$. This completes the proof.

\section{Proof of Lemma \ref{Lemma_Pmd_part1}} \label{Proof_Lemma_Pmd_part1}

By employing (\ref{Definition_Phi}) and (\ref{lemma_delta_constraint}), we can obtain
\begin{align} \notag
\Phi \left( {\frac{3}{2}\delta } \right) + \frac{1}{2}\delta  & = {\left( {2{D_U} + \Upsilon } \right)^{\frac{1}{2}}}{\left[ {\frac{{2{D_U} + \Upsilon }}{{{D_S}}}\left( {\frac{\Upsilon }{{{D_S}}} + 1} \right) + 2} \right]^{\frac{1}{2}}}\sqrt {\frac{3}{2}\delta }  + \frac{1}{2}\delta \\ \notag
& < \left\{ {{{\left( {2{D_U} + \Upsilon } \right)}^{\frac{1}{2}}}{{\left[ {\frac{{6{D_U} + 3\Upsilon }}{{2{D_S}}}\left( {\frac{\Upsilon }{{{D_S}}} + 1} \right) + 3} \right]}^{\frac{1}{2}}} + {\frac{1}{2}\Upsilon ^{\frac{1}{2}}}} \right\}\sqrt \delta  \\ \label{Phi_lambda}
& < \lambda,
\end{align}
and hence, $\Phi ( {\frac{3}{2}\delta } ) < \lambda$.

Furthermore, from (\ref{Define_tilde_D_j}), (\ref{Definition_lambda_j}) and (\ref{Definition_lambda}), we can obtain that 
\begin{align} \notag
\left| {{{\widetilde D}_j} - {D_j}} \right| & = {D_0}P_0^{\frac{1}{\gamma }}\left| {{{\left[ {{\tau _j} - F_j^{ - 1}\left( {{{\tilde p}_j}\left( {{\btheta _{\rm{T}}}} \right)} \right)} \right]}^{ - \frac{1}{\gamma }}} - {{\left[ {{\tau _j} - F_j^{ - 1}\left( {{p_j}\left( {{\btheta _{\rm{T}}}} \right)} \right)} \right]}^{ - \frac{1}{\gamma }}}} \right|\\ \label{D_j_difference_temp1}
& \ge {D_0}P_0^{\frac{1}{\gamma }}\mathop {\inf }\limits_{x \in \left[ {{\rho_{j}^{(\rm{L})}},{\rho_{j}^{(\rm{U})}}} \right]} \left| {\frac{{\partial {{\left[ {{\tau _j} - F_j^{ - 1}\left( x \right)} \right]}^{ - \frac{1}{\gamma }}}}}{{\partial x}}} \right|\left| {{{\tilde p}_j}\left( {{\btheta _{\rm{T}}}} \right) - {p_j}\left( {{\btheta _{\rm{T}}}} \right)} \right|\\  \notag 
& = {D_0}P_0^{\frac{1}{\gamma }}\mathop {\inf }\limits_{x \in \left[ {{\rho_{j}^{(\rm{L})}},{\rho_{j}^{(\rm{U})}}} \right]} \left| {\frac{{{{\left[ {{\tau _j} - F_j^{ - 1}\left( x \right)} \right]}^{ - \frac{{\gamma  + 1}}{\gamma }}}}}{{{f_j}\left( F_j^{ - 1}\left( x \right) \right)}}} \right|\left| {{{\tilde p}_j}\left( {{\btheta _{\rm{T}}}} \right) - {p_j}\left( {{\btheta _{\rm{T}}}} \right)} \right| \\  \label{tD_D_j_larger_temp1}
&  \ge \frac{\kappa {{D_0}P_0^{\frac{1}{\gamma }}{{\left[ {{\tau _j} - F_j^{ - 1}\left( {{\rho_{j}^{(\rm{L})}}} \right)} \right]}^{ - \frac{{\gamma  + 1}}{\gamma }}}}}{{\mathop {\sup }\limits_{x \in \left[ {{F_j^{-1}(\rho_{j}^{(\rm{L})})},{F_j^{-1}(\rho_{j}^{(\rm{U})})}}  \right]} {f_j}\left( x \right)}} \\  \label{tD_D_j_larger}
& >\lambda,
\end{align}
where (\ref{D_j_difference_temp1}) is due to (\ref{p_j_range}) and (\ref{tp_j_range}), and (\ref{tD_D_j_larger_temp1}) is from (\ref{Psi_j_larger}). 
Thus, we know
\begin{equation}
\cC \lp \btheta_j, \tD_j \rp \cap \cB\lp \btheta_\teT, \Phi\lp \frac{3}{2}\delta \rp \rp = \emptyset,
\end{equation}
since $\btheta_\teT \in \cC \lp \btheta_j, D_j \rp $ and $\Phi ( {\frac{3}{2}\delta } ) < \lambda$.

\begin{figure}
	\centerline{
		\includegraphics[height=0.5\textwidth]{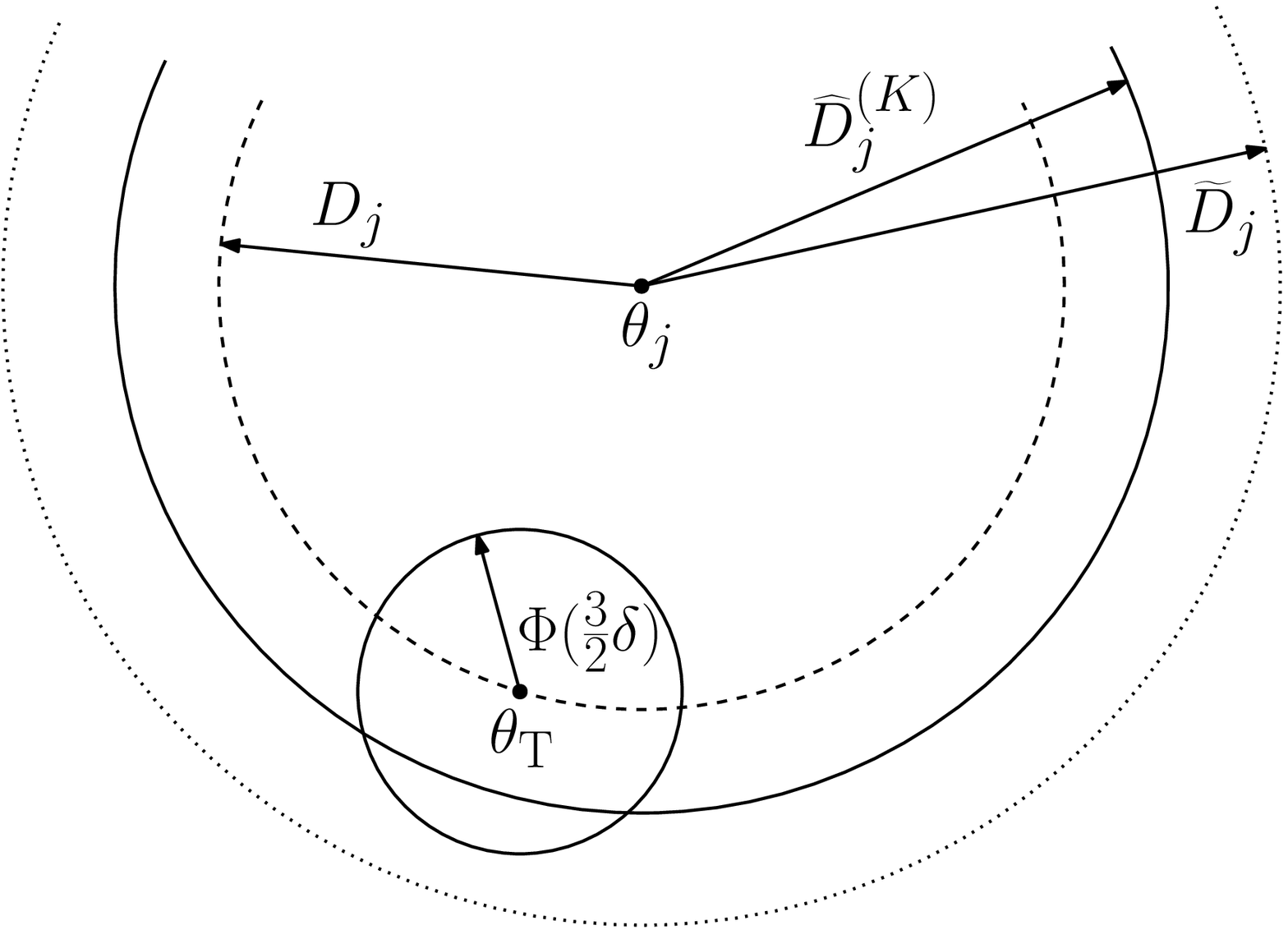}
	}
	\caption{Geometric illustration of (\ref{proof_lemma_temp1}).
	}	
	\label{Fig_Geometric_Illustration_temp1}	
\end{figure}

As illustrated in Fig. \ref{Fig_Geometric_Illustration_temp1}, if
\begin{equation}
{{\cal C}\left( {{\btheta _j},\hD_j^{(K)}} \right) \cap {\cal B}\left( {{\btheta _{\rm{T}}},\Phi \left( {\frac{3}{2}\delta } \right)} \right) \ne \emptyset },
\end{equation}
then
\begin{equation} \label{proof_lemma_temp1}
\left| {\hD_j^{(K)} - {{\widetilde D}_j}} \right| \ge  \left| {  {{\widetilde D}_j} - D_j   } \right| -  \Phi \left( {\frac{3}{2}\delta }  \right),
\end{equation}
which implies
\begin{equation}
\left| {\hD_j^{(K)} - {{\widetilde D}_j}} \right| \ge \lambda - \Phi \left( {\frac{3}{2}\delta }  \right) > \frac{1}{2}\delta,
\end{equation}
by employing (\ref{Phi_lambda}) and (\ref{tD_D_j_larger}).
Therefore,
\begin{equation}
{\bbP_1}\left( {{\cal C}\left( {{\btheta _j},\hD_j^{(K)}} \right) \cap {\cal B}\left( {{\btheta _{\rm{T}}},\Phi \left( {\frac{3}{2}\delta } \right)} \right) \ne \emptyset } \right)  \le {\bbP_1}\left( {\left| {\hD_j^{(K)} - {{\widetilde D}_j}} \right| \ge  \frac{1}{2}\delta } \right),
\end{equation}
which completes the proof.

\bibliographystyle{IEEEtran}
\bibliography{StochasticEncryption}

\end{document}